\documentclass[aps,prx,floatfix,superscriptaddress,twocolumn,tightenlines,nobibnotes,amsmath,amssymb]{revtex4}

\usepackage{graphicx}
\usepackage{amsmath}
\usepackage{amssymb}
\usepackage{epstopdf}
\usepackage{color}
\usepackage{natbib}
\usepackage{multirow}
\usepackage{array}
\usepackage{threeparttable}

\newcommand{\ket}[1] {|#1 \rangle}

\newcommand*{\rom}[1]{\expandafter\@slowromancap\romannumeral #1@}
\newcolumntype{C}[1]{>{\centering\let\newline\\\arraybackslash\hspace{0pt}}m{#1}}

\begin{document}

\title{Benchmarking high fidelity single-shot readout of semiconductor qubits}
\author{D. Keith}
\affiliation{Centre of Excellence for Quantum Computation and Communication Technology, School of Physics, University of New South Wales, Sydney, New South Wales 2052, Australia}
\author{S. K. Gorman}
\affiliation{Centre of Excellence for Quantum Computation and Communication Technology, School of Physics, University of New South Wales, Sydney, New South Wales 2052, Australia}
\author{L. Kranz}
\affiliation{Centre of Excellence for Quantum Computation and Communication Technology, School of Physics, University of New South Wales, Sydney, New South Wales 2052, Australia}
\author{Y. He}
\affiliation{Centre of Excellence for Quantum Computation and Communication Technology, School of Physics, University of New South Wales, Sydney, New South Wales 2052, Australia}
\author{J. G. Keizer}
\affiliation{Centre of Excellence for Quantum Computation and Communication Technology, School of Physics, University of New South Wales, Sydney, New South Wales 2052, Australia}
\author{M. A. Broome}
\altaffiliation{Current address: Department of Physics, University of Warwick, Coventry CV4 7AL, UK }
\affiliation{Centre of Excellence for Quantum Computation and Communication Technology, School of Physics, University of New South Wales, Sydney, New South Wales 2052, Australia}
\author{M. Y. Simmons}
\affiliation{Centre of Excellence for Quantum Computation and Communication Technology, School of Physics, University of New South Wales, Sydney, New South Wales 2052, Australia}
\date{\today}

\begin{abstract}
Determination of qubit initialisation and measurement fidelity is important for the overall performance of a quantum computer. However, the method by which it is calculated in semiconductor qubits varies between experiments. In this paper we present a full theoretical analysis of electronic single-shot readout and describe critical parameters to achieve high fidelity readout. In particular, we derive a model for energy selective state readout based on a charge detector response and examine how to optimise the fidelity by choosing correct experimental parameters. Although we focus on single electron spin readout, the theory presented can be applied to other electronic readout techniques in semiconductors that use a reservoir.
\end{abstract}

\maketitle

\section{Introduction}

Quantum computing relies on the preparation, control and measurement of quantum states~\cite{DiVincenzo2000}. In order to achieve scalable universal quantum computation the error rate of all these processes needs to be less than ${\sim}$1~\%---known as the fault-tolerant threshold for 2-dimensional surface codes~\cite{PhysRevA.83.020302,Fowler2012,Hille1500707,ogorman2016}. Recently, an emphasis has been placed on the quality of single and two-qubit gates through randomised benchmarking~\cite{PhysRevA.77.012307,PhysRevLett.106.180504,Barends2014,1367-2630-16-10-103032}. The usefulness of randomised benchmarking comes from removal of state preparation and measurement errors. It also scales polynomially with the number of qubits making it an efficient verification and validation method~\cite{Cross2016}. However, state preparation and measurement errors will also lower the overall fidelity of the quantum computer operation and always need to be considered for fault-tolerant quantum computation~\cite{PhysRevLett.113.220501}. Recent large-scale proposals for quantum computers in semiconductors utilise single electron or nuclear spins as the qubits~\cite{loss1998,kane1998,ogorman2016,Hille1500707,Tosi2017,Veldhorst2017}. The measurement of the electron spin in these architectures can be performed using a weakly coupled reservoir to the quantum dot/donor. Motivated by these proposals, we examine the fidelity of correctly identifying the spin of the electron using a nearby reservoir that is monitored by a charge sensor.

Over a decade ago, single-shot spin readout of an electron was first achieved in a semiconductor device~\cite{Elzerman2004} and since then it has been demonstrated using single donors~\cite{Morello2010} and donor quantum dots in semiconductors~\cite{Buch2013}, Si/SiGe quantum dots~\cite{PhysRevLett.106.156804}, Si-MOS quantum dots~\cite{VeldhorstM.2014} and nitrogen vacancy centres~\cite{Robledo2011}. The ability to perform high fidelity single-shot readout has  improved over the years reaching the point where single-shot spin readout can be performed above the fault-tolerant threshold~\cite{Watson2015}. However, the method used to determine the fidelity of single-shot readout has not been consistent between experiments making it difficult to directly compare one system with another. 

Independent of the system that is studied, the ability to distinguish between quantum states relies upon separating measurement distributions for each state using a particular threshold with respect to a signal from a detector. The measurement distributions are only dependent on a few experimentally accessible parameters including the noise spectrum, measurement bandwidth, temperature, magnetic field strength, tunnel rate to the reservoir, qubit energy separation, and the timing of the state conversion process. These factors have not always been consistently accounted for, and the fidelity analysis using energy selective spin readout has evolved since its first demonstration by Elzerman \textit{et al.} in 2004~\cite{Elzerman2004}. Elzerman \textit{et al.}~\cite{Elzerman2004} characterised the measurement fidelity of a single electron spin qubit by measuring the impact on detection errors from unwanted spin-flips, due to temperature or relaxation~\cite{PhysRevA.89.012313}, as well as charge sensor dark counts during readout. Since then the semiconductor electronic qubit field has mostly performed measurements and fidelity calculations utilising a peak filter to distinguish spin states, where the measured detector signal must cross a particular threshold value within a given readout time~\cite{PhysRevA.89.012313,Morello2010,Pla2012,VeldhorstM.2014,Watsone1602811,PhysRevApplied.8.034019}. 

Ref.~\cite{Morello2010} employed the now commonly used Monte-Carlo method to fit simulated signal histograms to experimental histograms to optimise the readout threshold and take post-processing errors into account. However, in this analysis the effect of the finite electron temperature and spin relaxation on the spin state during readout was not included. In the papers to follow, the effect of the finite electron temperature was determined using a variety of methods, such as electron spin resonance measurements~\cite{Pla2012,VeldhorstM.2014}, direct temperature measurements~\cite{Buch2013}, and tunnel rate measurements ~\cite{Watsone1602811,PhysRevApplied.8.034019}. The previously used Monte-Carlo method involves simulating the readout traces (generally with white Gaussian noise) based on the experimental parameters calculated from the experiment and comparing the resulting histograms to the experimental histograms. Other post-processing techniques have also been proposed such as wavelet-edge detection~\cite{0957-4484-26-21-215201}, Bayesian inference~\cite{PhysRevLett.111.126803} and maximum-likelihood estimation~\cite{PhysRevA.76.012325} to better detect the qubit states, yet these do not remove the need to fit to a histogram of numerically simulated readout traces.

It is important that an agreed upon methodology is used to both benchmark results and help optimise readout for future experimental work. Therefore, we propose a comprehensive, analytical approach with the capability to determine optimal thresholds for performing single-shot readout with the highest possible fidelity without relying on arbitrary numerics. The model we present is extendable to multi-qubit systems with regards to sequential readout and can be generalised to any noise spectrum. Specifically, we demonstrate its performance in the case of white Gaussian noise as an example, which is commonly used to model detector noise~\cite{Morello2010,Buch2013,Watsone1602811,PhysRevA.89.012313}. Using our model, we describe different limiting factors in readout fidelity, how to identify them and strategies to increase the fidelity once they have been identified. Finally, our method removes the need to rely on Monte-Carlo simulations~\cite{Morello2010} to calculate fidelities which we show in Supplementary Material I, can introduce a large error on the readout fidelity by inappropriate bin and simulation numbers, which are often not quoted.

\begin{figure}
\begin{center}
\includegraphics[width=1\columnwidth]{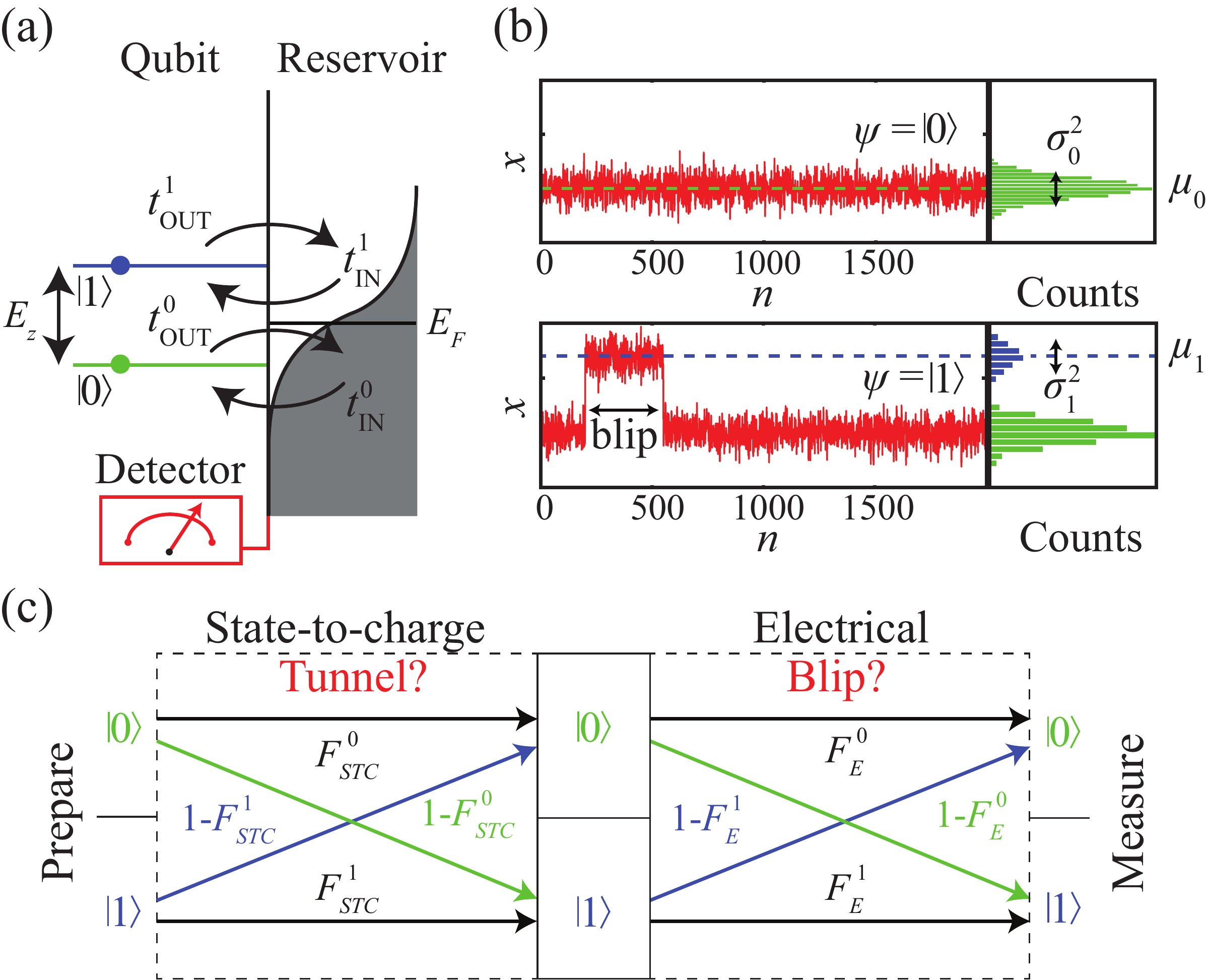}
\end{center}
\caption{{\bf Single-shot readout using a reservoir}. (a) Schematic of the Zeeman split energy levels ($\ket{0}$ and $\ket{1}$ separated by the Zeeman energy $E_Z$) in the system showing a charge detector that is used to determine if an electron has tunnelled between the qubit and reservoir. The electron tunnel rates are dictated by the Fermi broadening of the reservoir about the Fermi level $E_F$ according the the reservoir temperature. (b) Illustrative response from the detector for a low level, $\ket{0}$ and high level, $\ket{1}$. The `blip' in the detector response, $x$ indicates the measurement of $\ket{1}$. Right: Corresponding detector histograms showing the mean, $\mu_i$ and noise (variance), $\sigma_i^2$ for both $i{=}0, 1$. (c) Assignment of the qubit states based on the charge detector response. The qubit can be prepared in either the $\ket{0}$ or $\ket{1}$ spin state. Whether or not these states are preserved throughout spin-to-charge conversion depends on the conditional probability that the electron tunnelled ($F_{STC}^1$), or not ($F_{STC}^0$). Similarly, the qubit states are preserved throughout electrical readout depending on the conditional probability that a `blip' was successfully measured ($F_{E}^1$), or not ($F_{E}^0$). By following the arrows, the probability of correctly identifying the qubit state can be obtained.} 
\label{fig:intro}
\end{figure}

We outline our analytic approach in as general terms as possible, with sufficient detail, and based in experimental parameters to encourage applicability, consistency, and practicality. For the proceeding sections, we make no assumptions on the energy levels of the qubit and use the terminology of high level ($\ket{1}$), which is generally taken to be the excited state of the qubit and the low level ($\ket{0}$), usually assigned to the ground state as shown in Fig.~\ref{fig:intro}a. We also assume that the four tunnel times of the individual qubit states to a reservoir (or similar structure), $\{ t_{\textnormal{\tiny{IN}}}^0, t_{\textnormal{\tiny{OUT}}}^0, t_{\textnormal{\tiny{IN}}}^1, t_{\textnormal{\tiny{OUT}}}^1 \}$ are known. 

To determine the readout fidelity we must first understand the readout process itself. In the specific case of single-shot measurements of single electron spins \cite{Elzerman2004}, during the readout time, the reservoir Fermi level is tuned in between the Zeeman split electron spin states when in a global magnetic field, as in Fig.~\ref{fig:intro}a. This is done in such a way that only the excited spin state $\ket{1}$ possesses enough energy to tunnel to the reservoir, to then quickly be replaced by a ground state electron $\ket{0}$ tunnelling back from the reservoir. These tunnelling events are what cause the `blip' in the response of a nearby charge detector (see Fig.~\ref{fig:intro}b). During the qubit readout process the detector response, $x$ is monitored and the maximum value is recorded. We assume that the detector is monitored for a readout time, $t_r$  and sampled at a rate of $\Gamma_s$. The detection event is represented by a `blip' in the charge sensor response as a function of the number of samples $n$ made over time, $t$. The detection event corresponds to one of the two possible initial states ($\ket{0}$ or $\ket{1}$), which should ideally only occur when the electron is initially $\ket{1}$ to perform accurate readout. 

The errors in single-shot readout can come from either the conversion of quantum states to a measurable voltage or current signal or from post-processing where the qubit state can be incorrectly assigned from the measurement. The individual state fidelities $F_{i}$ are the conditional probabilities of correctly identifying the qubit states, $\ket{0}$ and $\ket{1}$, after the qubit is initially in the corresponding state. The state fidelity $F_0$ ($F_1$) can be broken down into fidelities related to different stages of the readout process as depicted in Fig.~\ref{fig:intro}c: the state-to-charge conversion (STC), which is related to $F_{STC}^0$ ($F_{STC}^1$) and electrical detection, related to $F_{E}^0$ ($F_{E}^1$) and are given by,
\begin{equation}
\label{eqn:f_0}
F_0 = F_{STC}^0 F_{E}^0 + (1 - F_{STC}^0)(1 - F_{E}^1),
\end{equation}
\begin{equation}
\label{eqn:f_1}
F_1 = F_{STC}^1 F_{E}^1 + (1 - F_{STC}^1)(1 - F_{E}^0),
\end{equation}
such that $F_{STC}^0 F_{E}^0$ is the conditional probability that the $\ket{0}$ qubit state did not tunnel to the reservoir and did not give a `blip' in the detector response. Similarly, $F_{STC}^1 F_{E}^1$ is the conditional probability that the $\ket{1}$ qubit state did tunnel to the reservoir and a `blip' was detected by the sensor. The second terms in Eq.~\ref{eqn:f_0} and Eq.~\ref{eqn:f_1} result from consecutive errors that cancel each other out. The overall measurement fidelity, $F_M$, which we define as the average detection fidelity of each qubit state is given by,
\begin{equation}
F_M = \frac{F_{0} + F_{1}}{2}
\end{equation}
We now detail the calculations of the STC and electrical detection fidelities required to calculate the overall measurement fidelity.

\section{State-to-charge conversion}

The STC process is used to maximise the probability (and hence, fidelity/visibility) that the detected `blip' is resultant from the $\ket{1}$ state, and not the $\ket{0}$ state, by optimising the readout time $t_r$. The STC visibility, $V_{STC}$ is dependent on $t^0_{\textnormal{\tiny{OUT}}}$, $t^1_{\textnormal{\tiny{OUT}}}$ and the $\ket{1}$ relaxation time, $T_1$,
\begin{equation}
V_{STC}(t_r; t^0_{\textnormal{\tiny{OUT}}}, t^1_{\textnormal{\tiny{OUT}}}, T_1) = F_{STC}^0 + F_{STC}^1 - 1,
\end{equation}
where the 0 level fidelity $F_{STC}^0$ is the probability that $\ket{0}$ does not tunnel to the reservoir due to thermal fluctuations and the 1 level fidelity $F_{STC}^1$ is the probability that $\ket{1}$ has tunnelled to the reservoir in a time, $t_r$. Assuming perfect electrical fidelity we can categorise the individual fidelities based on STC as per Fig.~\ref{fig:intro}c (left hand side),
{\renewcommand{\arraystretch}{1.5}
\begin{table}[h!]
\centering
\begin{tabular}{c c |  c  c }
\multicolumn{2}{c}{\multirow{2}{*}{ }} & \multicolumn{2}{c}{Measurement outcome}\\
& & No Tunnel & Tunnel\\\cline{2-4}
\parbox[t]{1mm}{\multirow{2}{*}{\rotatebox[origin=c]{90}{Qubit}}} \hspace{0.1cm} & $\ket{0}$ & $F_{STC}^0$ & $1 - F_{STC}^0$\\
& $\ket{1}$ & $1 - F_{STC}^1$ & $F_{STC}^1$
\end{tabular}
\end{table}

Errors occur when electron tunnelling occurs when the qubit is initially $\ket{0}$, or when there is no tunnelling resultant from the initial $\ket{1}$ state, caused mainly by the temperature of the reservoir and the energy separation of the two qubit states. High temperatures can thermally excite ground-state electrons to incorrectly tunnel, while high magnetic fields can increase the effects of relaxation. These give rise to two limiting situations for $V_{STC}$: readout time limited (RTL) and $T_1$ limited (TL).

Assuming we have detected a `blip' using the electrical threshold, there is some probability that it could be due to a $\ket{0}$ tunnelling out due to the finite temperature in the system. Determining this probability is the goal of the state-to-charge conversion analysis and was previously derived by Buch~\cite{Buch2013}; however, we repeat it here for completeness. We denote the state of the qubit by $\psi$ in the basis $\{\ket{0}, \ket{1}\}$. We define the initial time at the start of the readout phase as $t{=}0$, and describe the system dynamics using the rate equation,
\begin{equation}
\frac{d \psi}{d t} = \begin{pmatrix}
-\frac{1}{t^0_{\textnormal{OUT}}} & \frac{1}{T_1} \\
0 & -\frac{1}{t^1_{\textnormal{OUT}}} - \frac{1}{T_1}
\end{pmatrix} \psi,
\end{equation}
which has the solution,
\begin{multline}
\psi_0(t) = \frac{\psi_0(0) T_{\textnormal{OUT}}^2 + \psi_1(0) t^0_{\textnormal{OUT}} t^1_{\textnormal{OUT}}}{T_{\textnormal{OUT}}^2 e^{\frac{t}{t^0_{\textnormal{OUT}}}}} \\ - \frac{\psi_1(0) t^0_{\textnormal{OUT}} t^1_{\textnormal{OUT}}}{T_{\textnormal{OUT}}^2 e^{\frac{T_1 + t^1_{\textnormal{OUT}}}{t^1_{\textnormal{OUT}} T_1} t}},
\end{multline}
\begin{equation}
\psi_1(t) = \psi_1(0) e^{-\frac{T_1 + t^1_{\textnormal{OUT}}}{t^1_{\textnormal{OUT}} T_1} t},
\end{equation}
where, $\psi_i(0)$ is the initial probability of being in state $i$ and $T_{\textnormal{OUT}}^2{=}T_1 (t^0_{\textnormal{OUT}}{-}t^1_{\textnormal{OUT}}){+}t^0_{\textnormal{OUT}} t^1_{\textnormal{OUT}}$. The probability that a either qubit state generates a `blip' in the time, $t$, $N_{\textnormal{off}}(t)$ is,
\begin{align}
N_{\textnormal{off}}(t) &= 1 - \psi_0(t) - \psi_1(t) \\ &= F_{STC}^1 \psi_1(0) + (1 - F_{STC}^0) \psi_0(0),
\end{align}
and similarly, we label the probability that either qubit state does not generate a `blip' as,
\begin{align}
N_{\textnormal{on}}(t) &= \psi_0(t) + \psi_1(t) \\ &= (1 - F_{STC}^1) \psi_1(0) + F_{STC}^0 \psi_0(0).
\end{align}
From these two equations we can calculate the probability that the detected `blip' was from $\ket{1}$ and not from $\ket{0}$. First, we find $N_{\textnormal{on}}(t)$ when $\psi_0(0){=}1$ as this gives the probability that the $\ket{0}$ does not generate a `blip',
\begin{equation}
F_{STC}^0 = e^{-\frac{t}{t^0_{\textnormal{OUT}}}}.
\end{equation}
Similarly, for $N_{\textnormal{off}}(t)$ when $\psi_1(0){=}1$ we have
\begin{multline}
F_{STC}^1 = \frac{1}{T_{\textnormal{OUT}}^2}\Big[(1 - e^{-\frac{t}{t^0_{\textnormal{OUT}}}}) t^0_{\textnormal{OUT}} t^1_{\textnormal{OUT}} \\ + (e^{-\frac{T_1 + t^1_{\textnormal{OUT}}}{t^1_{\textnormal{OUT}} T_1} t} - 1) T_1 (t^1_{\textnormal{OUT}} - t^0_{\textnormal{OUT}})\Big],
\label{eqn:fstc1}
\end{multline}
which gives the probability of the $\ket{1}$ generating a `blip'. 

Now, to optimise the state-to-charge conversion we must find the maximum visibility between them. The visibility can be shown to be a single peaked function,
\begin{multline}
\label{eqn:v_stc}
V_{STC}(t) = \frac{T_1 (t^0_{\textnormal{\tiny{OUT}}} - t^1_{\textnormal{\tiny{OUT}}})}{T_{\textnormal{\tiny{OUT}}}^2} e^{-t\big(\frac{1}{T_1} + \frac{1}{t^0_{\textnormal{\tiny{OUT}}}} + \frac{1}{t^1_{\textnormal{\tiny{OUT}}}}\big)} \\
\times (e^\frac{t}{t^0_{\textnormal{\tiny{OUT}}}} - e^{-\frac{T_1 + t^1_{\textnormal{\tiny{OUT}}}}{t^1_{\textnormal{\tiny{OUT}}} T_1} t}).
\end{multline}
The optimum readout time $t_{opt}$ can then be found by differentiation with respect to $t$,
\begin{equation}
\label{eqn:topt}
t_{opt} = \frac{T_1 t^0_{\textnormal{\tiny{OUT}}} t^1_{\textnormal{\tiny{OUT}}}}{T_{\textnormal{\tiny{OUT}}}^2} \ln\Bigg(\frac{t^0_{\textnormal{\tiny{OUT}}} (T_1 + t^1_{\textnormal{\tiny{OUT}}})}{T_1 t^1_{\textnormal{\tiny{OUT}}}}\Bigg).
\end{equation}
This sets the optimal wait time for the readout step. 

The relative tunnel out times of the $\ket{0}$ and $\ket{1}$ level, assuming that $T_1{\gg}t^{i}_{\textnormal{OUT}}$, play a significant role in determining the readout fidelity. The tunnel rates are usually dependent on the Fermi distribution of the reservoir, therefore, we can find an approximation of their magnitude based on the temperature of the system. We assume $t^{i}_{\textnormal{OUT}}$ follows a Fermi distribution offset by the qubit energy separation, $E_Z$. That is, the tunnel rates are given by,
\begin{align}
\Gamma^{i}_{\textnormal{OUT}}(\epsilon) &= [1 - f(\epsilon \pm E_Z/2)] \Gamma_{\textnormal{OUT}},\\
\Gamma^{i}_{\textnormal{IN}}(\epsilon) &= f(\epsilon \pm E_Z/2) \Gamma_{\textnormal{IN}}
\end{align}
where,
\begin{equation}
f(\epsilon \pm E_Z/2) = \frac{1}{1 + e^{\frac{-(\epsilon \pm E_Z/2)}{k_B T}}},
\end{equation}
is the Fermi-Dirac function with $-$ for $\ket{0}$ and $+$ for $\ket{1}$, $\Gamma_{\textnormal{OUT}}$ ($\Gamma_{\textnormal{IN}}$) is the maximum tunnel rate out (in) of both qubit states, and $k_B T$ is the thermal energy of the system.  We can use two tunnel rates to find the detuning position about the reservoir, $\epsilon_{sr}$. Typically, $t^1_{\textnormal{OUT}}$ and $t^0_{\textnormal{IN}}$ are measured during the spin readout protocol. Assuming, $\Gamma_{\textnormal{OUT}}{=}\Gamma_{\textnormal{IN}}$, then the ratio between the two times is given by,
\begin{equation}
\frac{t^1_{\textnormal{OUT}}}{t^0_{\textnormal{IN}}} = \frac{f(\epsilon - E_Z/2)}{1 - f(\epsilon + E_Z/2)} = R_t,
\end{equation}
which gives,
\begin{equation}
R_t = \frac{1 + e^{\frac{E_Z - 2 \epsilon}{2 k_B T}}}{1 + e^{\frac{E_Z + 2 \epsilon}{2 k_B T}}},
\end{equation}
After rearranging and solving for $\epsilon$ (neglecting the imaginary solution), we have,
\begin{multline}
\epsilon_{sr} = k_B T \ln{\Bigg[\frac{e^{\frac{-Ez}{2 k_B T}}}{2 R_t} \Big(1 - R_t} \\ + \sqrt{(1 - R_t)^2 + 4 R_t e^{\frac{E_Z}{k_B T}}}\Big) \Bigg].
\end{multline}
This value of detuning can then be used to obtain the other two tunnel times in the system. For example, $t^0_{\textnormal{OUT}}$ can be found from using,
\begin{equation}
t^0_{\textnormal{OUT}} = \frac{1 - f(\epsilon_{sr} - E_Z/2)}{1 - f(\epsilon_{sr} + E_Z/2)} t^1_{\textnormal{OUT}} .
\end{equation}
Alternatively, if only one tunnel time is known, we can find the relative magnitude between two tunnel rates at $\epsilon{=}0$,
\begin{align}
\frac{\Gamma^1_{\textnormal{OUT}}}{\Gamma^0_{\textnormal{OUT}}} &= \frac{t^0_{\textnormal{OUT}}}{t^1_{\textnormal{OUT}}} =  e^{\frac{E_Z}{2 k_B T}} \\
t^0_{\textnormal{OUT}} &=  e^{\frac{E_Z}{2 k_B T}} t^1_{\textnormal{OUT}}.
\label{eqn:rates}
\end{align}

The ratio between the $t^{i}_{\textnormal{IN}}$ and $t^{i}_{\textnormal{OUT}}$ states can also be found by making similar assumptions. In this case,
\begin{align}
\frac{\Gamma^0_{\textnormal{IN}}}{\Gamma^0_{\textnormal{OUT}}} &= \frac{t^0_{\textnormal{OUT}}}{t^0_{\textnormal{IN}}} =  \frac{f(-E_Z/2)}{1 - f(-E_Z/2)} =  e^{\frac{E_Z}{2 k_B T}} \\ 
t^0_{\textnormal{OUT}} &= e^{\frac{E_Z}{2 k_B T}} t^0_{\textnormal{IN}},
\end{align}
and similarly,
\begin{align}
\frac{t^1_{\textnormal{OUT}}}{t^1_{\textnormal{IN}}} &=  \frac{f(E_Z/2)}{1 - f(E_Z/2)} = e^{- \frac{E_Z}{2 k_B T}} \\
t^1_{\textnormal{OUT}} &= e^{- \frac{E_Z}{2 k_B T}} t^1_{\textnormal{IN}}.
\end{align}
Eq.~\ref{eqn:v_stc} and Eq.~\ref{eqn:rates} allow for a relatively accurate estimate of the readout visibility for a particular temperature and the qubit energy separation. For spin qubits in a magnetic field $B$, the qubit energy separation is given by $E_Z=g\mu_BB$, where $g$ is the gyro-magnetic ratio and $\mu_B$ is the Bohr magneton. Hence, the tunnel rates which help determine $V_{STC}$ have an exponential dependence on both the reservoir temperature and magnetic field strength. The magnetic field also reduces the excited state relaxation time $T_1$ at large fields with a $T_1{\propto}B^{-\alpha}$ dependence \cite{Morello2010}. Having shown that $V_{STC}$ has a strong dependence on the magnetic field and reservoir temperature we now move on to discuss the readout visibility in each of the limiting cases caused by these key factors and how $V_{STC}$ can be optimised.

\begin{figure}
\includegraphics[width=1\columnwidth]{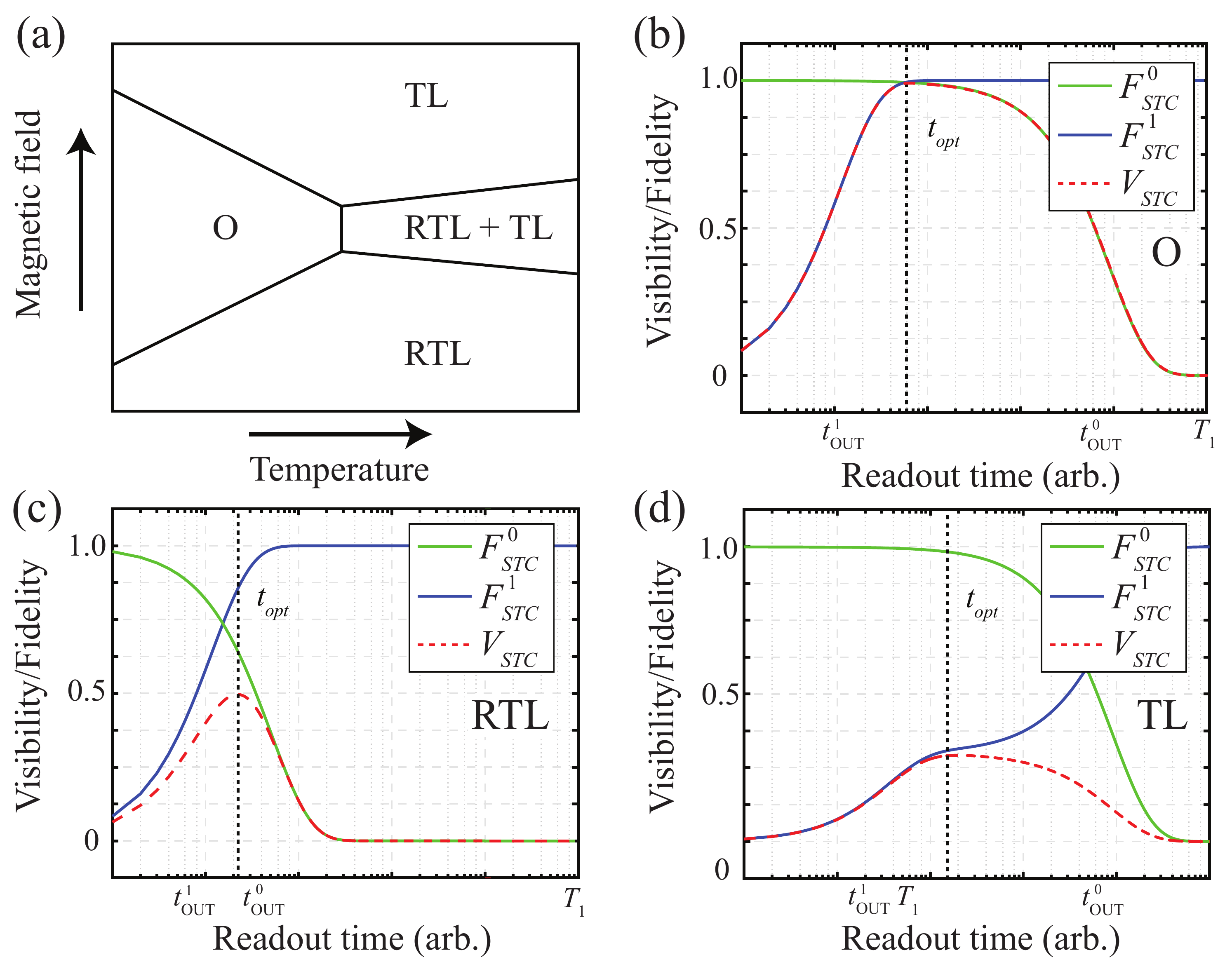}
\caption{{\bf Calculated state-to-charge conversion visibilities of optimal, readout time, and $T_1$ limited single-shot readout.} (a) Schematic of the state-to-charge conversion visibility, $V_{STC}$ as a function of the reservoir temperature and magnetic field for a single spin qubit showing where the fidelity is limited by different factors. (b) {\bf Optimal visibility (O)} The probability of $\ket{0}$ not tunnelling ($F_{STC}^0$, green) and $\ket{1}$ tunnelling ($F_{STC}^1$, blue) along with $V_{STC}$ (red) approaches 1 at the optimal readout time. The highest $V_{STC}$ is obtained for low temperatures and an energy separation that is large but not too large to substantially reduce the $T_1$ time (assuming $T_1{\propto}B^{-\alpha}$) \cite{Morello2010}. (c) {\bf Readout time limited (RTL)}. This limit applies to systems where $t^0_{\textnormal{\tiny{OUT}}}$, the time for the incorrect state to tunnel out is short or comparable to $t^1_{\textnormal{\tiny{OUT}}}$, thus limiting the optimal readout time. Only a fraction of $t^1_{\textnormal{\tiny{OUT}}}$ events will be captured within such limited readout window. (d) {\bf $T_1$ limited (TL)}. Short relaxation times of the high energy state increases the probability of relaxing to the lower energy state before it has a chance to tunnel out, thus no `blips' are registered by the detector.}
\label{fig:limitingcase_stc}
\end{figure}

\subsection{Optimal state-to-charge visibility}

In Fig.~\ref{fig:limitingcase_stc}b we plot the fidelities, $F_{STC}^0$ and $F_{STC}^1$ as well as $V_{STC}$ as a function of the readout time. Since $F_{STC}^0$ corresponds to the probability of $\ket{0}$ \emph{not} tunnelling out to the reservoir, at $t{=}0$, $F_{STC}^0{=}1$. The fidelity $F_{STC}^0$ then follows an exponential decay as $\ket{0}$ becomes more likely to tunnel off to the reservoir. The fidelity $F_{STC}^1$ represents the probability that the $\ket{1}$ state has tunnelled off to the reservoir. At $t{=}0$, $F_{STC}^1{=}0$ as there has not been a chance for the qubit state to tunnel to the reservoir. As $t{\rightarrow}\infty$, $F_{STC}^0{\rightarrow}0$ and $F_{STC}^1{\rightarrow}1$ as the probability that the qubit tunnels approaches unity. Therefore, there is an optimum readout time, $t_{opt}$ that maximises $V_{STC}(t)$ and offers the best compromise between $F_{STC}^0$ and $F_{STC}^1$. The state-to-charge conversion visibility follows the $F_{STC}^1$ curve for short $t$ and then follows $F_{STC}^0$ after the optimal readout time, $t_{opt}$. The best $V_{STC}$ is obtained for low reservoir temperatures and for a large qubit energy splitting since this maximises the ratio of $t^0_{\textnormal{\tiny{OUT}}}$ to $t^1_{\textnormal{\tiny{OUT}}}$. 

\subsection{Readout time limited}

The readout time limit occurs when the optimum readout time of the system calculated from the state-to-charge analysis is close to the individual tunnel out times, $t^0_{\textnormal{\tiny{OUT}}}$ and $t^1_{\textnormal{\tiny{OUT}}}$ (Fig.~\ref{fig:limitingcase_stc}c). This will occur when the difference between $t^0_{\textnormal{\tiny{OUT}}}$ and $t^1_{\textnormal{\tiny{OUT}}}$ becomes very small, that is $t^1_{\textnormal{\tiny{OUT}}}/t^0_{\textnormal{\tiny{OUT}}}{\rightarrow}1$ making it difficult to find an optimal readout time, hence decreasing both $F_{STC}^0$ and $F_{STC}^1$. This is generally an indication that the state levels are not sufficiently separated in energy compared to the temperature broadening of the reservoir. If the relative tunnel rates cannot be changed then this scenario is extremely difficult to overcome since it means the temperature in the system needs to be reduced. We determine that for $F_M{>}99$~\% then $t^1_{\textnormal{\tiny{OUT}}}{/}t^0_{\textnormal{\tiny{OUT}}}{\gtrsim}800$, which for a qubit energy splitting, $E_Z$ corresponds to $E_Z{/}k_B T{\approx}13$. For electron spins in silicon, this corresponds to magnetic field to temperature ratio of $B{/}T{\approx}10$~T/K.

\subsection{$T_1$ limited}

Finally, the last situation is when the $T_1$ of the $\ket{1}$ level is close to the optimal readout time calculated from the state-to-charge analysis. A distinct plateau in the state-to-charge visibility can be observed that limits $F_{STC}^1$, shown in Fig.~\ref{fig:limitingcase_stc}d. This is due to the large fraction of $\ket{1}$ states relaxing to $\ket{0}$ and not causing a `blip' in the charge sensor response. We find that provided $T_1{\gtrsim}100t^1_{\textnormal{\tiny{OUT}}}$ then $F_M$ can be above $99$\%. Again, this situation is difficult to overcome without the ability to change the relative tunnel rates.

\section{Electrical readout}

\begin{figure}[t!]
\includegraphics[width=1\columnwidth]{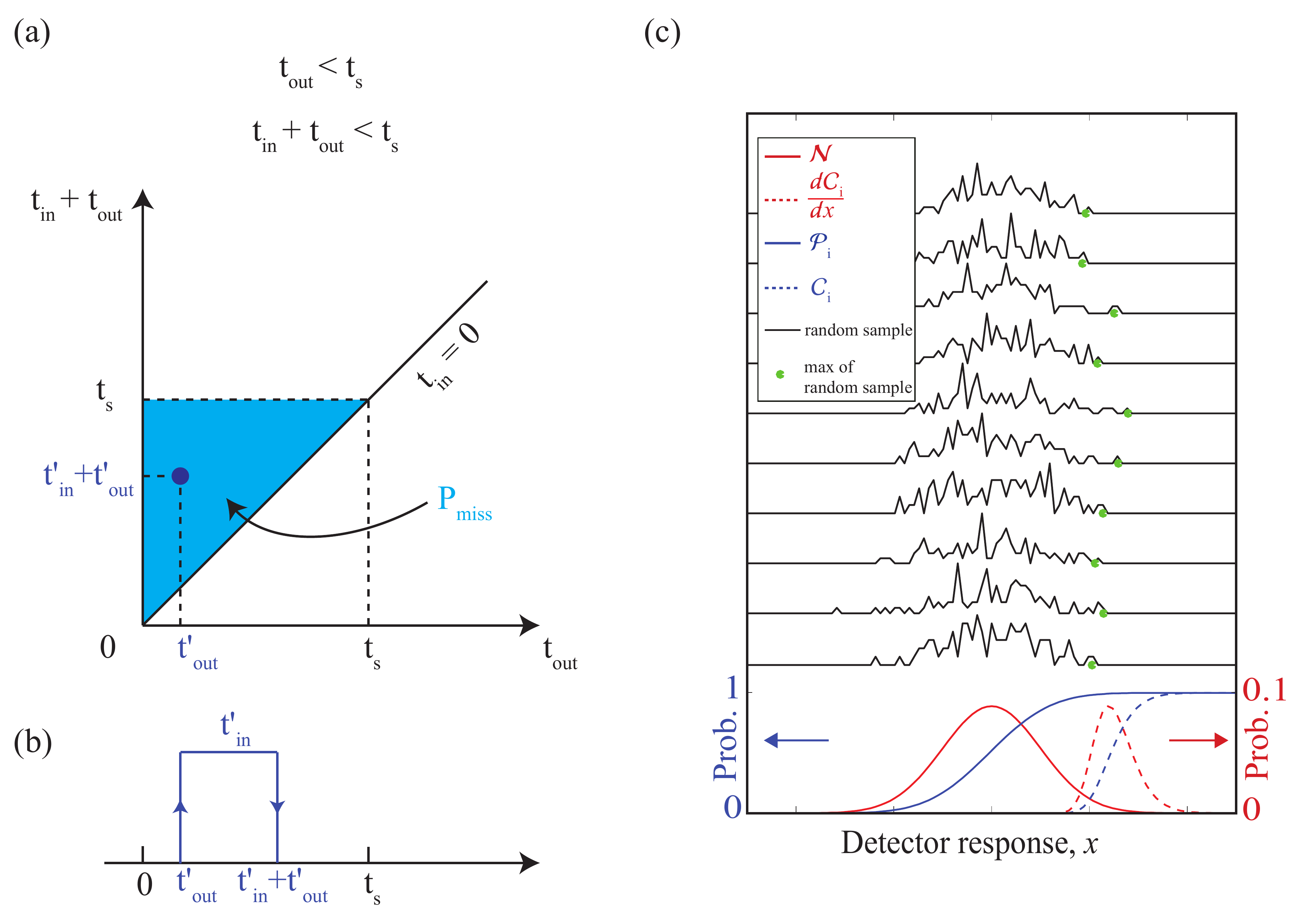}
\caption{{\bf Electrical readout probabilities} (a) `Blips' in the detector response can be described using $t^1_{\textnormal{OUT}}$ and $t^0_{\textnormal{IN}}+t^1_{\textnormal{OUT}}$. By definition, for a `blip' to physically exist, $t^0_{\textnormal{IN}}>0$. If $t^0_{\textnormal{IN}}+t^1_{\textnormal{OUT}}{\leq}t_s$ as well, then the `blip' will be undetectable in the detector response, and hence contribute to $P_{miss}$. The points that fulfill these criteria are depicted by the blue shaded region. The shaded area is equivalent to an area with height $t^0_{\textnormal{IN}}+t^1_{\textnormal{OUT}}=t_s$ and width $t^0_{\textnormal{IN}}=t_s/2$. b) Depiction of an example `blip' that would be undetectable and contribute to $P_{miss}$. The time until the initial edge of the `blip' is given by $t^1_{\textnormal{OUT}}$, and the duration of the `blip' is given by $t^0_{\textnormal{IN}}$. c) Distribution histograms (solid black lines) of traces randomly sampled from a Gaussian PDF (solid red line) with corresponding CDF (solid blue line). Also included is the corresponding PDF (red dashed line) and CDF (blue dashed line) if only the maximum (green dots) of each trace is sampled. Sampling each maximum point inherently skews both the PDF and CDF to higher detector response values in a way that can be calculated analytically for the purpose of calculating electrical readout fidelities if the initial PDF ($\mathcal{N}$) is known.}
\label{fig:diagrams}
\end{figure}

The electrical visibility $V_{E}$ is the ability of the charge detector to resolve the `blip' in the detector response $x$ and is related to the sample rate of the charge sensor, $\Gamma_s$, $t_{\textnormal{\tiny{OUT}}}^1$, $t_{\textnormal{\tiny{IN}}}^0$, and the sensitivity index, $D'$ defined as
\begin{equation}
D'=\frac{\mu_1{-}\mu_0}{\sqrt{\frac{1}{2}(\sigma_1^2{+}\sigma_0^2)}},
\end{equation}
where, $\mu_{i}$ ($\sigma_{i}$) is the mean (standard deviation) of the $i{=}0,1$ levels of the charge sensor response. In the case when $\sigma_0=\sigma_1$, $D'$ is equivalent to the signal-to-noise ratio. Lastly, $V_{E}$ also depends on the filter cut-off frequency, $f_c$ used to filter the charge sensor response,
\begin{equation}
V_{E}(x; D', \Gamma_s, t_{\textnormal{\tiny{OUT}}}^1, t_{\textnormal{\tiny{IN}}}^0, f_c) = F_{E}^0 + F_{E}^1 - 1,
\label{eqn:ten}
\end{equation}
where $F_{E}^0$ is the probability of $\ket{0}$ not causing a `blip' above a threshold value $x$ and $F_{E}^1$ is the probability of $\ket{1}$ generating a `blip' (assuming that the $V_{STC}{=}1$) in the charge detector response as per Fig.~\ref{fig:intro}c (right hand side),
{\renewcommand{\arraystretch}{1.5}
\begin{table}[h!]
\centering
\begin{tabular}{c c |  c  c }
\multicolumn{2}{c}{\multirow{2}{*}{ }} & \multicolumn{2}{c}{Measurement outcome}\\
& & No `blip' & `blip'\\\cline{2-4}
\parbox[t]{1mm}{\multirow{2}{*}{\rotatebox[origin=c]{90}{Qubit}}} \hspace{0.1cm} & $\ket{0}$ & $F_{E}^0$ & $1 - F_{E}^0$\\
& $\ket{1}$ & $1 - F_{E}^1$ & $F_{E}^1$
\end{tabular}
\end{table}

Errors arise when a `blip' occurs within the readout time when the qubit is initially $\ket{0}$, or when there is no `blip' within the readout time when the initial state is $\ket{1}$. The key factors that cause these errors, and hence reduce $V_{E}$, include a sample rate too slow to detect fast `blips', high noise that disguises a potential `blip', or filtering such that fast `blips' are removed from the measured signal. 

Optimisation of the electrical fidelity requires finding the threshold that gives the maximum ability to distinguish between $0$ and $1$ in the detector response. Therefore, we must calculate the value of the detector response, $x$ over the duration of $t_r$ which maximises $V_E$, as it will be used as the optimal threshold value $x_{opt}$ to distinguish between the two states.

First we want to find the probability $P_{miss}$ of missing a `blip' due to the finite sample time of the detector. The tunnel out event of the $\ket{1}$ state can occur anywhere within the interval $\delta t{=}t_s{-}\tau$ where $\tau$ is some point in time less than the sample time, $t_s{=}1/\Gamma_s$. Therefore, the probability of detecting a high level `blip' is a sum between the exponential distribution of $\ket{1}$ normalised over the interval $0$ to $t_s$,
\begin{equation}
p_1(t) = \frac{e^{-t/t^1_{\textnormal{OUT}}}}{t^1_{\textnormal{OUT}} (1 - e^{-t_s/t^1_{\textnormal{OUT}}})}
\end{equation}
and the distribution of $\ket{0}$,
\begin{equation}
p_0(t) = \frac{e^{-t/t^0_{\textnormal{IN}}}}{t^0_{\textnormal{IN}}},
\end{equation}
such that,
\begin{align}
p_{det} &= \int_0^t p_1(t - \tau) \cdot p_0(\tau) d\tau \\ &= \int_0^t \frac{e^{-(t - \tau)/t^1_{\textnormal{OUT}}}}{t^1_{\textnormal{OUT}} (1 - e^{-t_s/t^1_{\textnormal{OUT}}})} \frac{e^{-\tau/t^0_{\textnormal{IN}}}}{t^0_{\textnormal{IN}}} d\tau.
\end{align}
This convolution gives the probability of detecting a `blip' of length, $t$,
\begin{equation}
p_{det} = \Big(\frac{e^{R^1_s}}{e^{R^1_s} - 1}\Big)\frac{e^{-t/t^0_{\textnormal{IN}}} - e^{-t/t^1_{\textnormal{OUT}}}}{t^0_{\textnormal{IN}} - t^1_{\textnormal{OUT}}},
\end{equation}
where $R^1_s{=}t_s/t^1_{\textnormal{OUT}}$. Next, we find the total probability of missing all `blips', $P_{miss}$. We are interested in the time $t^1_{\textnormal{OUT}}{+}t^0_{\textnormal{IN}}{\leq}t_s$ for $t^0_{\textnormal{IN}}>0$ which, for simplicity, is equivalent to $t^1_{\textnormal{OUT}}{\leq}t_s{/}2$ as depicted by Fig.~\ref{fig:diagrams}a, hence, 

\begin{equation}
P_{miss} = 1 - \int^{t_s/2}_0 p_{det} dt,
\end{equation}
which results in,
\begin{equation}
\label{eq:pmiss}
P_{miss} = 1 - \frac{(1 - e^{(R^1_s - R^0_s)/2}) R^1_s}{(1 - e^{R^1_s/2})(R^1_s - R^0_s)},
\end{equation}
where $R^0_s{=}t_s/t^0_{\textnormal{IN}}$. All the `blips' with values of $t^1_{\textnormal{OUT}}$ and $t^0_{\textnormal{IN}}$ that fall within the blue shaded region of Fig.~\ref{fig:diagrams}a contribute to $P_{miss}$, and have the general shape of the example shown in Fig.~\ref{fig:diagrams}b. 

\begin{figure*}[t!]
\includegraphics[width=1\textwidth]{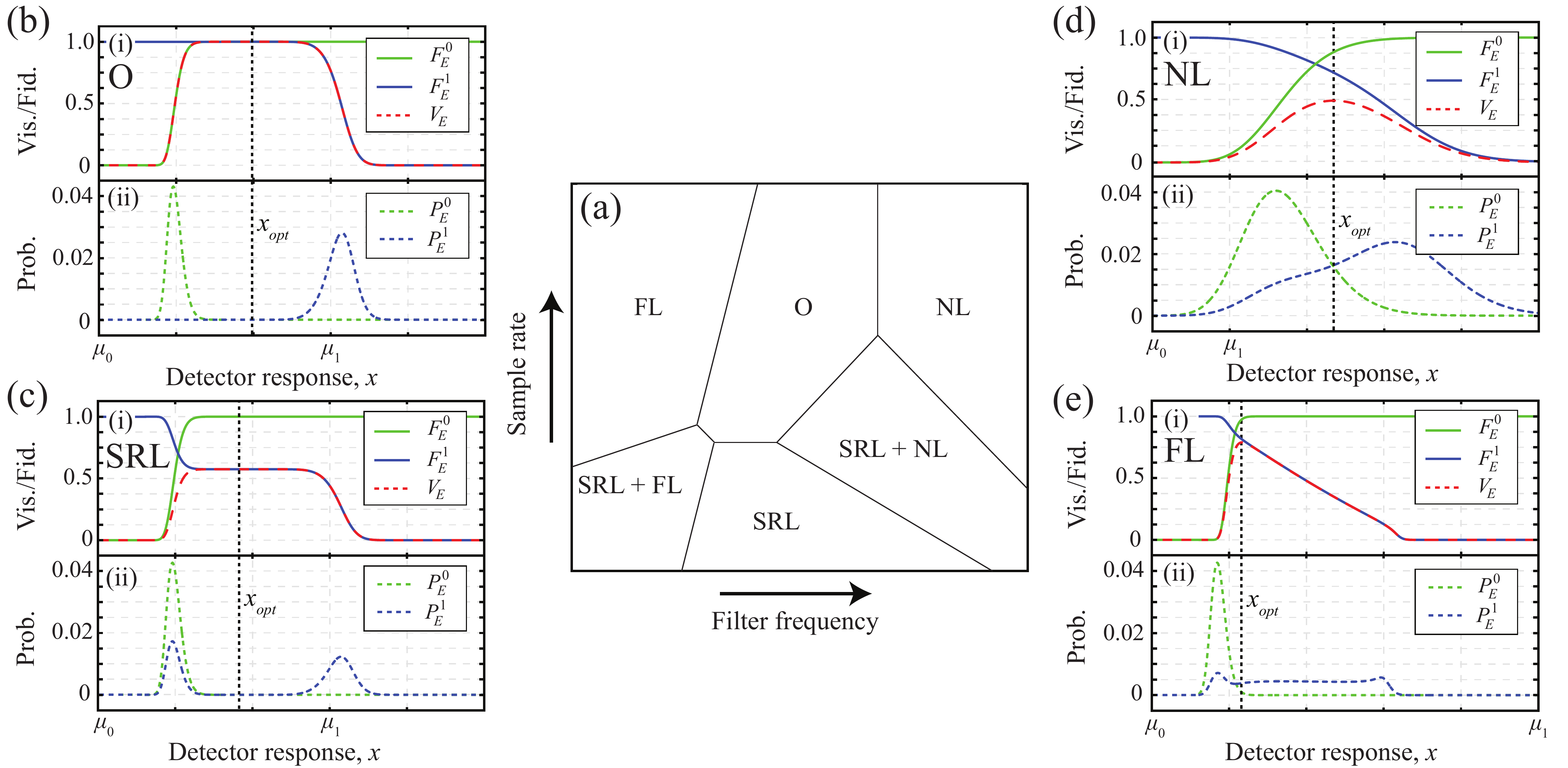}
\caption{{\bf Calculated electrical visibilities for optimal, sample rate, noise, and filter limited single-shot readout.} (a) Schematic of the electrical visibility as a function of the detector sample rate and filter cut-off frequency showing where the fidelity is limited by different factors. (b) {\bf Optimal visibility (O)}. The green, $F_{E}^0$ (blue, $F_{E}^1$) lines represent the fidelities of $\ket{0}$ ($\ket{1}$). The readout distributions of detecting $\ket{0}$ (green, $P_{E}^0$) and $\ket{1}$ (blue, $P_{E}^1$) is shown in subplots (ii). The optimal readout position is obtained with a high sample rate and careful optimisation of the filter cut-off frequency. In (b)(i) electrical visibility approaches 1 in-between the two distributions of (b)(ii) indicating a complete separation of the two readout distributions. (c) {\bf Sample rate limited (SRL)}. The fast tunnelling events happen in-between the sampling of the detector, causing a high number of false $\ket{0}$ counts seen as a $P_{E}^1$ peak underneath $P_{E}^0$ in (c)(ii). (d) {\bf Noise limited (NL)}. The signal-to-noise ratio is very low, thus the electrical fluctuations on the sensor trigger the false high-state counts. The readout distributions of $0$ and $1$ are overlapping, which limits the total visibility. (e) {\bf Filter limited (FL)}. In this scenario, the filter cut-off frequency is faster than the characteristic tunnelling time of the `blips', $t_{\textnormal{\tiny{IN}}}^0$. This gives rise to a number of false $\ket{0}$ counts. In this case the electrical visibility has characteristic asymmetrical shape shown in (d)(i). The readout distribution of $1$ has a long tail that extends down to the peak of $P_{E}^0$ due to the low Bessel filter frequency.}
\label{fig:limitingcase_ele}
\end{figure*}

Assuming we know the type and magnitude of the noise and average levels of the two states $\mu_i$, we can write the individual readout state fidelities as a function of the detector response,
\begin{equation}
F_{E}^0 = \mathcal{C}_0(x),
\label{gamma}
\end{equation}
\begin{equation}
F_{E}^1 = (1 - P_{miss}) [1 - \mathcal{C}_1(x)] + P_{miss} [1 - \mathcal{C}_0(x)],
\label{delta}
\end{equation}
where $\mathcal{C}_i(x)$ is the cumulative distribution function (CDF) for the maximum of the individual $0$ and $1$ readout distributions. For clarity, we depict several examples of histograms of detector response traces sampled from a Gaussian probability distribution function (PDF). The corresponding CDF then describes the probability of sampling a point below a given threshold. Fig.~\ref{fig:diagrams}c shows how both the PDF ($\frac{d \mathcal{C}_1(x)}{dx}$) and CDF ($\mathcal{C}_1(x)$) are skewed towards positive $x$ when only sampling the maximum of each trace as is the case during electrical readout.

Here, $F_{E}^0$ and $F_{E}^1$ are the CDFs of the measured readout traces and correspond to the fidelity of distinguishing whether a `blip' occurred between the readout distributions $P_{E}^i$, which we define as,
\begin{equation}
P_{E}^i = \bigg{|}\frac{dF_{E}^i}{dx}\bigg{|}.
\end{equation}
The missed `blips' due to the finite readout time are taken into account in the STC. The electrical visibility that distinguishes between the two qubit levels, from combining Eqs. \ref{eqn:ten}, \ref{gamma} and \ref{delta}, is then given by,
\begin{equation}
V_E(x) = (1 - P_{miss})[\mathcal{C}_0(x) - \mathcal{C}_1(x)].
\label{eq:ve}
\end{equation}
We can immediately see the importance of $P_{miss}$ on the readout fidelity as it limits the maximum $V_E(x)$. The optimum threshold, $x_{opt}$ is the value of $x$ that maximises $V_E(x)$. This can be found by differentiating Eq.~\ref{eq:ve} and setting the condition,
\begin{equation}
\frac{d V_E(x)}{dx} = (1 - P_{miss})\bigg{[}\frac{d \mathcal{C}_0(x)}{dx} - \frac{d \mathcal{C}_1(x)}{dx}\bigg{]} = 0,
\end{equation}
which corresponds to the value of $x$ where the two partial readout distributions are equal, $d \mathcal{C}_0(x){/}dx{=}d \mathcal{C}_1(x){/}dx$. We can now use these equations to investigate various limiting cases of the system. Similarly to STC, we now describe the optimal scenario (not limited by any particular experimental parameter). We also show the schematic phase diagram for the different limiting cases for $V_E(x)$ in Fig.~\ref{fig:limitingcase_ele}a. There are four distinct regions, optimal (O), sample rate limited (SRL), noise limited (NL), and filter limited (FL) which we will discuss in the following sections.

\subsection{Optimal electrical visibility}

In Fig.~\ref{fig:limitingcase_ele}b(i) we plot the fidelities, $F_{E}^0$ and $F_{E}^1$ as well as the electrical visibility, $V_E(x){=}F_{E}^0{+}F_{E}^1{-}1$ as a function of the detector response threshold, $x$. The $F_{E}^0$ level corresponds to the probability that the maximum value of the readout trace of $\ket{0}$ will be \emph{less} than $x$. Therefore, in Fig.~\ref{fig:limitingcase_ele}b(i) the probability, $F_E^0$ begins at zero for small $x$ and increases with a skewed Gaussian distribution as $x$ increases until it reaches 1, indicating that the distribution will always be less than those values of $x$ where $F_{E}^0{=}1$. For $F_{E}^1$ the condition is reversed; that is, $F_{E}^1$ corresponds to the probability that the maximum value of the readout trace of $\ket{1}$ will be \emph{greater} than $x$. For small $x$, $F_{E}^1{=}1$ indicating that the readout trace will always have a maximum value above $x$. As $x$ increases the probability that the maximum value lies above $x$ decreases and eventually there will never be a maximum value of the readout trace that is above $x$ which corresponds to $F_{E}^1{=}0$. The optimal threshold value, $x_{opt}$ can vary dramatically and its exact position will depend on what factor is limiting $V_E(x)$.

The lower panel in Fig.~\ref{fig:limitingcase_ele}b(ii) shows the distributions $P_{E}^0$ (green) and $P_{E}^1$ (blue) that can be used to visualise the difference between the two measured levels in the readout trace. For high fidelity readout, these two distributions do not overlap and are well separated in $x$.

\subsection{Sample rate limited}

The first limiting case we consider is when the sample rate of the detector is too low to be able to detect fast tunnelling events. This is characterised by a flat plateau region in the electrical fidelity in Fig.~\ref{fig:limitingcase_ele}c(i) at value below $V_E(x){=}1$. The sample rate does not affect the state-to-charge conversion fidelity and hence it can be arbitrarily high. The individual readout distributions are clearly distinguishable; however, there is a large number of events in the $P_E^1$ state that lie under $P_E^0$ shown in Fig.~\ref{fig:limitingcase_ele}c(ii). These events are faster than $\Gamma_s$ and are not measured by the detector, hence reduce $F_{E}^1$. Note that since the tunnelling events are stochastic there will always be a finite number of events faster than $\Gamma_s$. This limiting case can be easily remedied by increasing the sample rate of the sensor. Using Eq.~\ref{eq:pmiss} for $P_{miss}$, we find that the required sample rate for $F_M{>}99$~\% is $\Gamma_s{\gtrsim}12{/}t^0_{\textnormal{\tiny{IN}}}$.

\subsection{Noise limited}

If the detector is not sufficient filtered or has poor noise characteristics then the ability to distinguish between the two levels becomes difficult. When this is the case, the electrical visibility is limited by the noise of the charge detector. Noise limited electrical fidelities are characterised by an almost symmetrical peak in $V_E(x)$ (Fig.~\ref{fig:limitingcase_ele}d(i)) where the two readout distributions clearly overlap with each other, see Fig.~\ref{fig:limitingcase_ele}d(ii). This makes it difficult to optimise the detector response threshold, and reduces both $F_{E}^0$ and $F_{E}^1$. The noise limited scenario is more difficult to overcome compared to the sample rate limited situation. We can optimise the charge sensor using low-noise amplifiers~\cite{HoEomDayLeDucEtAl2012,doi:10.1063/1.4941421} or adjust the filter frequency to limit some of the noise in the device. However, reducing the filter frequency can also have a detrimental effect on the readout fidelity as high frequency `blips' can also be attenuated. Assuming white Gaussian noise, we find that it is possible that $F_M{>}99$~\% can be achieved with a sensitivity index $D'$ as low as 3 provided $\Gamma_s$ is sufficiently fast to resolve the `blips' in the readout trace. Note that this $D'$ is based on the integration time and sample rate rather than the readout time which is required to be longer due to the stochastic tunnelling processes.

\subsection{Filter limited}

The detector is normally low pass filtered in readout experiments to remove high frequency noise from the readout trace. However, this also filters the high frequency `blips' which reduces $F_{E}^1$, as well as the overall electrical visibility. In the case of a filter limited charge detector the peak in the electrical visibility will be asymmetrical with the $0$ distribution being quite sharp compared to the $1$ level as shown in Fig.~\ref{fig:limitingcase_ele}e(i). This is due to $P_E^1$ exhibiting an extremely long tail extending towards $P_E^0$ (Fig.~\ref{fig:limitingcase_ele}e(ii)). This scenario can be readily fixed by increasing the filter cut-off frequency. However, as the filter frequency is increased more noise couples into the charge sensor. Therefore, there is a trade off between the filter limited and noise limited scenario. The filter limit is much easier to improve and should essentially be increased until the noise in the system begins to dominate the electrical fidelity, that is, when the peak becomes symmetrical as in Fig.~\ref{fig:limitingcase_ele}d(i).

\section{Applications and discussion}

In this section we describe a number of applications and extensions of the model presented in the paper.

\subsection{Initialisation fidelity}

The initialisation fidelity can be found using a similar method to the STC visibility calculation. The rate equation model in the basis $\{\ket{z}, \ket{0}, \ket{1}\}$ where $\ket{z}$ is the state when the qubit is emptied is given by,
\begin{equation}
\frac{d \psi}{d t} = \begin{pmatrix}
-\frac{1}{t^0_{\textnormal{IN}}} - \frac{1}{t^1_{\textnormal{IN}}} & \frac{1}{t^0_{\textnormal{OUT}}} & \frac{1}{t^1_{\textnormal{OUT}}} \\
\frac{1}{t^0_{\textnormal{IN}}} & -\frac{1}{t^0_{\textnormal{OUT}}} & \frac{1}{T_1} \\
\frac{1}{t^1_{\textnormal{IN}}} & 0 & -\frac{1}{t^1_{\textnormal{OUT}}} - \frac{1}{T_1}
\end{pmatrix} \psi,
\label{eq:initfidelity}
\end{equation}
and assume the system starts in this state, $\psi_z(0){=}1$. The solution to this system of equations can be found analytically; however, the solutions are rather unwieldy. Instead, to get an approximation for the ideal initialisation time, we assume the electron cannot tunnel back to the reservoir, that is $t^0_{\textnormal{OUT}}{=}t^1_{\textnormal{OUT}}{\rightarrow}\infty$. It is worth noting that having a short $T_1$ aids in initialisation as the $\ket{1}$ is expected to decay quickly to the $\ket{0}$ state even if it were accidentally loaded. Therefore, we are interested in the regime where $t^{i}_{\textnormal{IN}}{\leq}T_1$.

The solution to the system of equations when $\psi_z(0){=}1$ is given by,
\begin{multline}
\psi_0(t) = \frac{e^{- \frac{t (1 + T_{t} T_1)}{T_1}}}{T^2_{\textnormal{IN}}} \Big[ t^0_{\textnormal{IN}} T_1 (e^{\frac{t (1 + T_{t} T_1)}{T_1}} - e^{t T_{t}}) \\
 + t^0_{\textnormal{IN}} t^1_{\textnormal{IN}} (e^{\frac{t}{T_1}} - e^{\frac{t (1 + T_{t} T_1)}{T_1}}) \\
 + t^1_{\textnormal{IN}} T_1 (e^{\frac{t (1 + T_{t} T_1)}{T_1}} - e^{\frac{t}{T_1}}) \Big],
\end{multline}
\begin{equation}
\psi_1(t) = \frac{t^0_{\textnormal{IN}} e^{- \frac{t (1 + T_{t} T_1)}{T_1}} ( e^{t T_{t}} - e^{\frac{t}{T_1}} )}{T^2_{\textnormal{IN}}},
\end{equation}
\begin{equation}
\psi_z(t) = e^{-t T_{t}},
\end{equation}
where, $T^2_{\textnormal{IN}}{=}T_1 (t^1_{\textnormal{IN}}{+}t^0_{\textnormal{IN}}){-}t^0_{\textnormal{IN}} t^1_{\textnormal{IN}}$ and
\begin{equation}
T_t = \frac{t^1_{\textnormal{IN}} + t^0_{\textnormal{IN}}}{t^1_{\textnormal{IN}} t^0_{\textnormal{IN}}}.
\end{equation}
Finally, we define the initialisation fidelity as,
\begin{equation}
F_I = \psi_0(t).
\end{equation}
The optimum $F_I$ is more difficult to define compared to $F_M$ because of the influence of $T_1$. Therefore, we define the optimal $F_I$ that maximises $F_I$ while also minimising the initialisation time, $t_I$. This time corresponds to the maximum chance of loading a $\ket{1}$ state. Therefore, $t_I$ can be calculated by differentiation of $\psi_1(t)$,
\begin{equation}
t_I = \frac{t^1_{\textnormal{IN}} t^0_{\textnormal{IN}} T_1}{T^2_{\textnormal{IN}}} \ln{(T_1 T_t)}.
\label{eq:initialtimeapprox}
\end{equation}
This time represents the time where the $T_1$ process starts dominating the initialisation process. That is, where the majority of the $\ket{0}$ are due to relaxation of the $\ket{1}$ state. Therefore, it represents a minimal time that maximises the initialisation fidelity.

\begin{figure}
\begin{center}
\includegraphics[width=1\columnwidth]{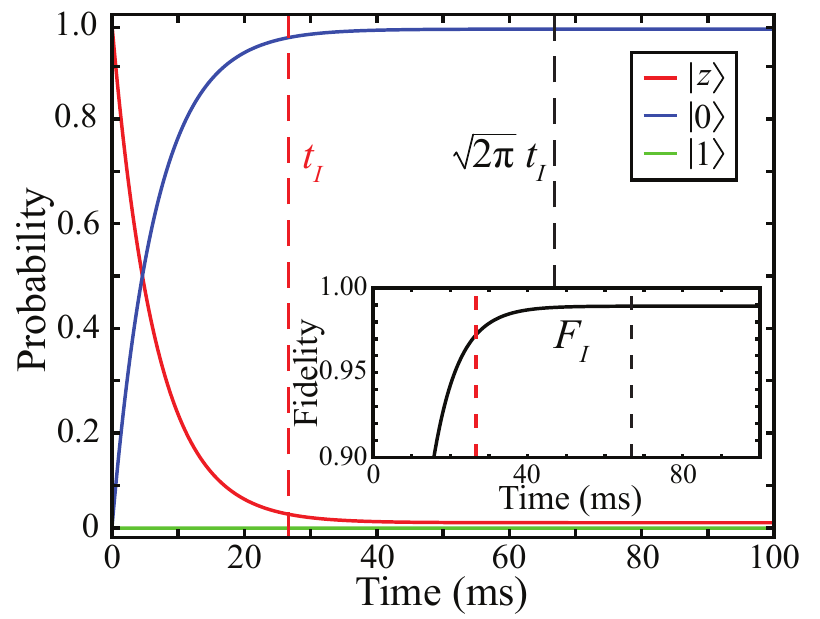}
\end{center}
\caption{{\bf Calculation of the initialisation fidelity and optimisation of the initialisation time}. Probability of the three qubit states during initialisation calculated using Eq.~\ref{eq:initfidelity} with the parameters from Broome($L$). The system begins in the empty state, $\ket{z}$ and then $\ket{0}$ state quickly becomes populated due to tunnelling from the reservoir to the qubit. The $\ket{1}$ states remains almost completely unpopulated due to the long tunnel in time, $t^1_{\textnormal{IN}}{\approx}2$ s. Two estimates for the initialisation time are shown by the dotted lines, $t_I$ (red) and $\sqrt{2 \pi} t_I$ (black). The latter of these two times is more conservative and has an initialisation fidelity, $F_I{=}98.9$ \%. The inset shows the initialisation fidelity as a function of initialisation time.}
\label{fig:initialisationfidelity}
\end{figure}

To demonstrate the initialisation fidelity calculation, in Fig.~\ref{fig:initialisationfidelity} we show the solution to Eq.~\ref{eq:initfidelity} for the data of Broome($L$) \cite{broome}. We initialise the system in the empty state and then watch how the qubit states are populated as a function of time. The $\ket{0}$ state quickly becomes populated due to direct tunnelling from the reservoir to the qubit. The $\ket{1}$ state becomes slightly populated during the initial tunnelling period $t{<}20$ ms; however, due to the slow tunnel rate into the $\ket{1}$ state from the reservoir there is never any significant population. We then show two different initialisation times, $t_I$ calculated from Eq.~\ref{eq:initialtimeapprox} and $\sqrt{2 \pi} t_I$. The first time, $t_I{=}26$ ms gives an initialisation fidelity, $F_I{=}97.2$ \%. The second initialisation time estimate, $\sqrt{2 \pi} t_I{=}65$ ms is more conservative and gives an initialisation fidelity, $F_I{=}98.9$ \%. The factor $\sqrt{2 \pi}$ was chosen based on examining a number of different initialisation fidelity calculations and shows a good compromise between initialisation time and fidelity. To obtain the actual initialisation fidelity (and time) the full system should be used to calculate $\psi(t)$. However, the above analysis offers a simple estimate to set the initialisation time.

\subsection{Calculation of $V_E$ assuming white Gaussian noise}

White Gaussian noise assumes that the levels, $\ket{0}$ and $\ket{1}$ have a Gaussian noise distribution with a constant spectral density at all frequencies. In this section, we find $\mathcal{C}_0$ and $\mathcal{C}_1$ for this type of noise.

We denote the mean levels of $\ket{0}$ and $\ket{1}$ in the $x$ domain as $\mu_0$ and $\mu_1$. Both $\ket{0}$ and $\ket{1}$ have a Gaussian noise distribution centred about their mean, $\mu_i$,
\begin{equation}
\mathcal{N}_i(x;\mu_i,\sigma_i^2) = \frac{1}{\sqrt{2 \pi} \sigma_i}e^{-\frac{(x - \mu_i)^2}{2 \sigma_i^2}}.
\end{equation}
The levels have an associated noise $\sigma_i^2$ and we define the sensitivity index,
\begin{equation}
D' = \frac{\mu_1 - \mu_0}{\sqrt{\frac{1}{2} (\sigma^2_1 + \sigma^2_0)}},
\end{equation}
which reduces to the signal-to-noise ratio (SNR) if $\sigma_1{=}\sigma_0$.

During the readout process, we are interested in the maximum value of the detector response during the readout time. Therefore, we want to build our state distributions by taking the maximum of $\mathcal{N}(x;\mu_i,\sigma_i^2)$ over a single readout trace.

First, we will consider the lower level, $\ket{0}$. For a fixed sample number of the readout trace $n_r{=}t_r/t_s$, the CDF for the maximum of a sampled Gaussian is simply the product of $n_r$ individual Gaussian distributions. Therefore, $\mathcal{C}_0(x)$ is given by,
\begin{equation}
\mathcal{C}_0(x) = \mathcal{P}_0(x)^{n_r},
\label{eq:CL}
\end{equation}
where $\mathcal{P}_i(x)$ is the CDF of a single Gaussian, given by,
\begin{equation}
\mathcal{P}_i(x) = \frac{1}{2} \Big[ 1 + \textnormal{erf} \Big( \frac{x - \mu_i}{\sqrt{2} \sigma_i} \Big) \Big].
\end{equation}

The $\ket{1}$ level CDF is more complicated since it involves a combination of both $\ket{0}$ and $\ket{1}$. The tunnelling events of $\ket{1}$ follow an exponential distribution in time, which can be defined as,
\begin{equation}
\mathcal{E}(n, n_i, n_{\textnormal{max}}) = \frac{e^{\frac{1 - n}{n_i}}}{n_i (1 - e^{\frac{1 - n_{\textnormal{max}}}{n_i}})},
\end{equation}
where $n{=}t/t_s$ is the sample number in the readout trace, $n_{\textnormal{max}}$ is the maximum number of samples in the distribution, and $n_i{=}t^{i}_{\textnormal{OUT}/\textnormal{IN}}/t_s$ is the characteristic length of $\ket{1}/\ket{0}$ in the readout trace. The $\mathcal{C}_1(x)$ can then be calculated by weighting the $\ket{0}$ and $\ket{1}$ probability distributions by $\mathcal{E}(n)$,
\begin{multline}
\mathcal{C}_1(x) = \int_{s = 1}^{n_r - 1} \mathcal{E}(s, n_0, n_r - 1) \times \\ \Bigg[ \int_{n = 1}^{n_r - s} \eta(n) \mathcal{S}_n dn
+ \int_{n = n_r - s}^{\infty} \eta(n) \mathcal{S}_{n_r - s} dn \Bigg] ds,
\label{eq:CH}
\end{multline}
where,
\begin{equation}
\mathcal{S}_n = \bigg(\frac{n}{n_r} \mathcal{P}_1(x) + \frac{n_r - n}{n_r} \mathcal{P}_0(x)\bigg)^{n_r},
\label{eq:sn}
\end{equation}
and
\begin{equation}
\eta(n) = \frac{e^{(1 - n)/n_1}}{n_1}.
\end{equation}
Here, the integration is carried out over two different scenarios. The first integral in the brackets describes the `blips' of a length, $n_r{-}s$, where $s$ is the length of the readout trace before the `blip'. These `blips' are fully resolved during the readout. The second integral describes `blips' that are actually longer than the $n_r{-}s$ and therefore become artificially shortened to exactly $n_r{-}s$. $\mathcal{S}_n$ is the relative probability of the $\ket{0}$ and $\ket{1}$ CDFs over the readout trace weighted by $\eta(n)$ over the entire readout trace. Eq.~\ref{eq:CH} can be numerically integrated to obtain $\mathcal{C}_1(x)$.

At this time, we introduce the effect of filters on the readout trace. The filter in readout experiments is usually low-pass and is characterised by a cut-off frequency, $f_c$ which we define as the -3 dB attenuation in the gain amplitude factor, $\mathcal{G}(f,f_c)$. The noise, $\sigma_i^2$ is then given by,
\begin{equation}
\sigma_i^2(f_c) = 2 A_n^2 f_c,
\label{eq:An}
\end{equation}
where $A_n$ is the noise power spectral density in units of $x/\sqrt{\textnormal{Hz}}$. This value can be found experimentally by simply measuring $\sigma_i^2(f_c)$ at a known $f_c$ and inverting Eq.~\ref{eq:An}. 

Finally, the filter also attenuates the amplitude of the `blip' in the readout trace since it has some frequency components above $f_c$. To account for this, we convert the cut-off frequency of the filter into the inverse number domain, $m{=}1/n$ of the readout trace,
\begin{equation}
m_c = \frac{f_c}{\Gamma_s},
\end{equation}
and apply it to $\mu_1$,
\begin{equation}
\mu_1(n) = \max{[h_p(n, m_c)]}(\mu_1 - \mu_0) + \mu_0,
\end{equation}
where $h_p(n, m_c)$ is the pulse response of the filter with cut-off, $m_c$. Therefore, any $\ket{1}$ levels with a frequency number much less than $m_c$ will be limited to $\mu_0$. These new noise, $\sigma_i^2(f)$ and mean, $\mu_1(n)$ parameters then need to be included in $\mathcal{P}_i(x)$ in Eqs.~\ref{eq:CL} and \ref{eq:CH}. The values of $\sigma_i^2(f)$ and $\mu_1(n)$ will be different depending on the filter used in the experimental setup. However, in Supplementary Materials II we give the calculation for an 8th order Bessel filter commonly used in experiments.

\subsection{An example of the optimisation of experimental parameters}

\begin{figure}
\includegraphics[width=1\columnwidth]{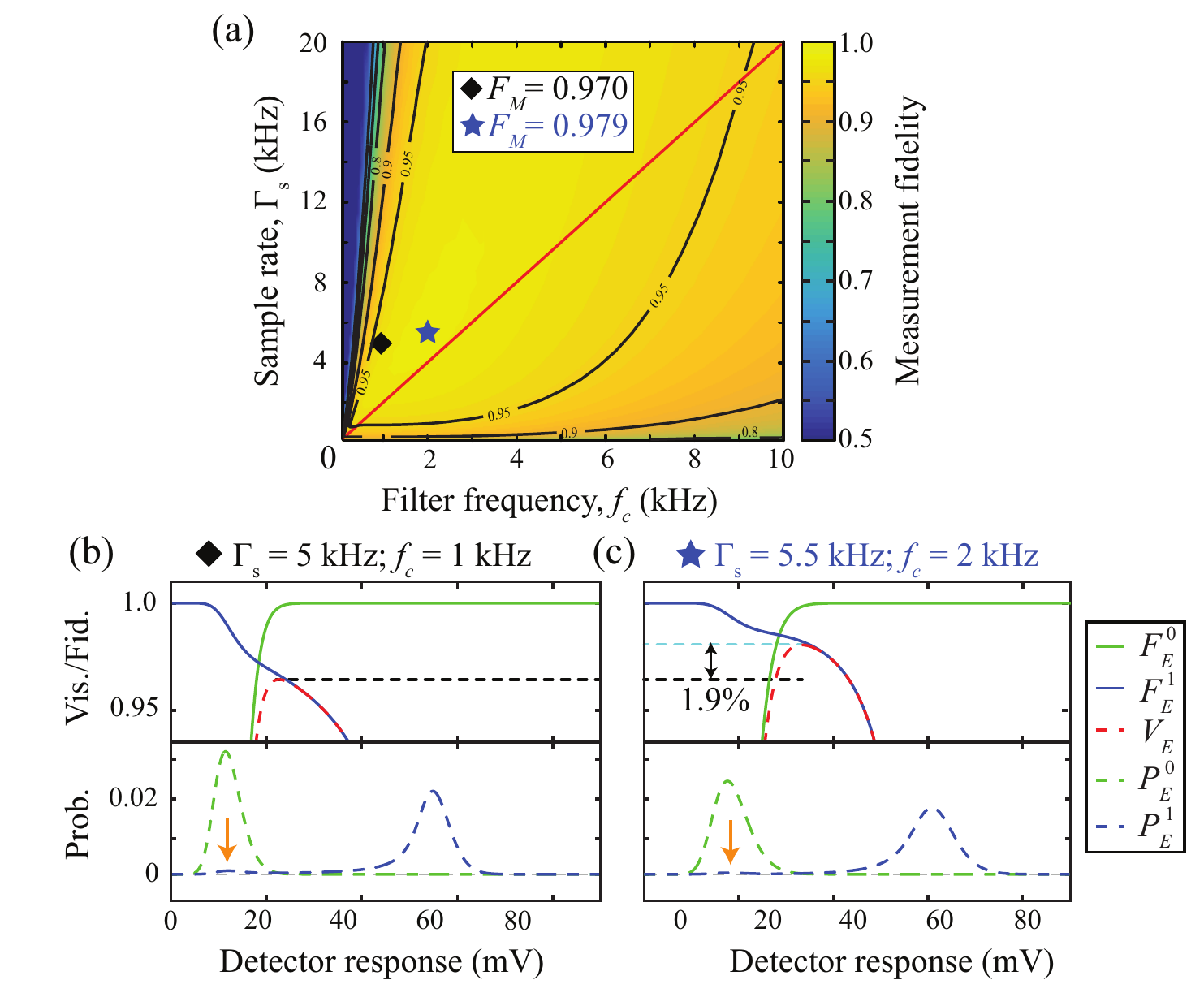}
\caption{{\bf Optimising the sample rate and filter frequency for single-shot spin readout.} (a) The measurement fidelity, $F_M$ as a function of sample rate, $\Gamma_s$ and filter frequency, $f_c$ for an electron spin qubit on a 2P donor dot (Broome($L$))~\cite{broome}. The black diamond is $F_M$ for $\Gamma_s{=}5$~kHz and $f_c{=}1$~kHz used in the experiment. The blue star is the optimum fidelity over the interval investigated ($\Gamma_s{=}5.5$~kHz and $f_c{=}2$~kHz). The red line corresponds to $\Gamma_s{=}2f_c$, the sample rate usually assumed to be correct based on the Shannon-Nyquist sampling theorem~\cite{shannon98}, which falls below the optimal fidelity point. (b) The electrical visibility and state distributions as a function of the detector response for $\Gamma_s{=}5$~kHz and $f_c{=}1$~kHz used in the experiment (black diamond). There is a significant number of $1$ states that lie underneath the $0$ state distribution (orange arrow). There is also a large tail on the $1$ state distribution indicating that the measurement fidelity is limited by a combination of sample rate and filter frequency. (c) The electrical visibility and state distributions as a function of the detector response for $\Gamma_s{=}5.5$~kHz and $f_c{=}2$~kHz (blue star). Using these parameters there is a clear change (1.9~\%) in the optimal $V_E$ caused by fewer missed $1$ states underneath the $0$ state readout distribution (orange arrow). This amounts to an increase in $F_M$ by $0.9$~\%.}
\label{fig:broomecompare}
\end{figure}

\begin{table*}[t]
\begin{threeparttable}
\caption{Comparison of reported fidelities and those calculated using the model in this paper with the same reported readout time $t_{rep}$ and detector response threshold $x_{rep}$. We observe that the state-to-charge visibilities $V_{STC}$ have a large impact on the overall measurement fidelity, thus should always be taken into account.}
\label{table2}
\centering
\begin{tabular}{  p{2.6cm} | C{2.0cm}  C{1.5cm} C{1.5cm} C{1.5cm} C{1.5cm} |  C{1.5cm}  C{1.5cm}  C{1.5cm} }
	\hline
	\hline
	\multicolumn{1}{c|}{\multirow{2}{*}{Reference}}& \multicolumn{5}{c|}{Reported Values} & \multicolumn{3}{c}{Calculated Values}  \\ 
	\multicolumn{1}{c|}{} 	& $t_{rep}$ 		& $x_{rep}$ 		& $V_{STC}$ (\%) 		& $V_E$ (\%) 			& $F_M$ (\%) 			& $V_{STC}$ (\%) 		& $V_E$ (\%) 			& $F_M$ (\%)  		\\
	\hline
		\cite{Elzerman2004} Elzerman  				& $0.5$ ms		& $0.73$ nA		& N/A			& $65.0$			& $82.5$			& $79.9 \pm 1.8$ 			& $52.7 \pm 0.2$			& $71.0 \pm 0.6$		\\
	\cite{Morello2010} Morello  				& $100$ $\mu$s  	& $1.1$ nA 		& N/A 			& $92.0$ 			& $96.0$ 			& $100$ 			& $74.7$ 			& $87.3$ 		\\
		\cite{PhysRevLett.106.156804} Simmons  		& $200$ ms 	& N/A 		& N/A 			& $88.0$ 			& $94.0$ 			& $97.8 \pm 0.3$ 			& N/A 			& N/A \\
		\cite{Nowack} Nowack(R)\tnote{a} 			& $2$ ms			& $220$ pA\tnote{b} & N/A			& N/A			& $86 \pm 1$			& $77.1 \pm 1.8$ 			& $94.8$ 			& $86.5 \pm 0.9$		\\
		\cite{Pla2012} Pla  					& $1$ ms			& $370$ pA		& $72 \pm 1$			& $82 \pm 2$			& $77 \pm 2$ 		& $40.1$ 			& $89.4$			& $67.9$		\\
	\cite{Buch2013} Buch  				& $40$ ms 		& $5.4$ pA		& $97.0$ 			& $96.2$ 			& $96.5$ 			& $96.1$ 			& $92.9$ 			& $94.6$  	\\
		\cite{VeldhorstM.2014} Veldhorst  			& $1$ ms\tnote{b} & N/A  & $96.3$ 			& $87.3$ 			& $92.0$ 			& $95.7$ 			& N/A  		& N/A 	\\
	 \cite{Watson2015} Watson($D^0$)\tnote{a} 		& $55 \pm 0.05$ ms 	& $120$ pA 		& $99.6$ 			& $99.6$ 			& $99.6$ 			& $99.6 \pm 0.2$ 			& $99.4$ 			& $99.5 \pm 0.1$ 		\\
	\cite{Watson2015} Watson($D^-$)\tnote{a}  		& $1 \pm 0.005$ ms 	& $1.2$ nA 		& $99.5$ 			& $97.4$ 			& $98.4$ 			& $99.2 \pm 0.1$ 			& $96.6 \pm 0.5$ 			& $97.9 \pm 0.3$ \\
		\cite{Watsone1602811} Watson($D1$)\tnote{a}  		& $58$ ms 	& $170$ pA\tnote{b} 		& $99.9$ 			& $99.8$ 			& $99.8$ 			& $99.9$ 			& $98.4 \pm 0.1$ 			& $99.2$ 		\\
	\cite{Watsone1602811} Watson($D2$)\tnote{a} 		& $62$ ms 	& $170$ pA\tnote{b} 		& $99.9$ 			& $99.7$ 			& $99.8$ 			& $99.9$ 			& $98.3 \pm 0.1$ 			& $99.1$ \\
		\cite{broome} Broome(L)\tnote{a} 			& $10.5 \pm 0.1$ ms & $22 \pm 1$ mV 	& $97.9 \pm 0.1$ 	& $94.6 \pm 1.0$ 	& $96.2 \pm 1.1$ 	& $97.9{\pm}0.5$ 			& $96.2{\pm}0.1$ 			& $97.1{\pm}0.3$ 	\\
	\cite{broome} Broome(R)\tnote{a}  			& $209 \pm 30$ ms	& $26 \pm 2$ mV  	& $98.7 \pm 0.2$ 	& $96.5 \pm 2.0$ 	& $97.6 \pm 2.1$ 	& $98.7{\pm}0.6$ 			& $96.5{\pm}0.1$ 			& $97.6{\pm}0.3$ \\
	\hline
	\hline
	
\end{tabular}
\begin{tablenotes}
\item[a] Names within parenthesis are taken from initial reference to distinguish between readout performed on different quantum dots/transitions.
\item[b] Value estimated from the figures appearing in a given reference
\end{tablenotes}
\end{threeparttable}
\end{table*}

\begin{table*}[t]
\begin{threeparttable}
\caption{Readout fidelities calculated using the analytic model presented here while using the original reported experimental parameters (see Supplementary Material III) with optimised values for the readout time $t_{opt}$ and detector response threshold $x_{opt}$. This optimisation improved fidelities up to 8\% compared to fidelities calculated with reported thresholds (Gain is equal to optimised $F_M$ minus the calculated $F_M$ from Table~\ref{table2}).}
\label{table3}
\centering
\begin{tabular}{  p{2.6cm} | C{2cm}  C{2cm} C{1.5cm}  C{1.5cm}  C{1.5cm} | C{1.5cm} }
	\hline
	\hline
	\multicolumn{1}{c|}{\multirow{2}{*}{Reference}}& \multicolumn{5}{c|}{Optimised Values}& \multicolumn{1}{c}{ } \\ 
	\multicolumn{1}{c|}{} 	& $t_{opt}$ 		& $x_{opt}$ 		& $V_{STC}$ (\%)		& $V_E$ (\%) 			& $F_M$ (\%) 			& Gain (\%) 		\\
	\hline
		\cite{Elzerman2004} Elzerman  &$0.46 \pm 0.01$ ms	&$0.74 \pm 0.01$	&$79.9 \pm 1.8$	&$67.6 \pm 0.2$	&$75.8 \pm 0.6$ 	& $+4.8$	\\
		\cite{Morello2010} Morello  &$175 \mu$s 	&$1.52$ nA	&$100.0$	&$92.4$	&$96.2$	&$+8.9$	\\
		\cite{PhysRevLett.106.156804} Simmons  	&$139 \pm 7$ ms 	&$2118$ pA	&$98.1 \pm 0.3$	&$92.5 \pm 0.1$	&$95.4 \pm 0.2$	& N/A	 \\
		\cite{Nowack} Nowack(R)\tnote{a} 	&$1.65 \pm 0.04$ ms 	&$260$ pA &$77.6 \pm 1.8$	&$97.2$	&$87.7 \pm 0.9$	&$+1.2$	\\
		\cite{Pla2012} Pla  &$0.55$ ms 	&$646$ pA	&$47.7$	&$92.9$	&$72.2$	&$+4.3$		\\
		\cite{Buch2013} Buch  	&$22$ ms	&$7.2$ pA &$97.4$	&$94.2$	&$95.9$	&$+1.3$	\\
		\cite{VeldhorstM.2014} Veldhorst  &$0.15$ ms 	&$344$ pA	&$99.2$	&$91.6$	&$95.4$	& N/A		\\
	 	\cite{Watson2015} Watson($D^0$)\tnote{a} &$53.4 \pm 5$ ms 	&$241$ pA	&$99.6 \pm 0.2$	&$99.4$	&$99.5 \pm 0.1$	&$0.0$		\\
		\cite{Watson2015} Watson($D^-$)\tnote{a}  &$0.98 \pm 0.06$ ms 	&$1.32$ nA	&$99.2 \pm 0.1$	&$97.1 \pm 0.5$	&$98.2 \pm 0.3$	&$+0.3$	 \\
		\cite{Watsone1602811} Watson($D1$)\tnote{a} &$58.5 \pm 2.6$ ms 	&$188 \pm 1$ pA	&$99.9$	&$99.5 \pm 0.1$	&$99.7$	&$+0.5$		\\
		\cite{Watsone1602811} Watson($D2$)\tnote{a} &$57.4 \pm 3$ ms 	&$187 \pm 1$ pA	&$99.9$	&$99.3 \pm 0.1$	&$99.6$	&$+0.5$		 \\
		\cite{broome} Broome(L)\tnote{a} 	&$10.6 \pm 0.2$ ms 	&$22.8 \pm 0.2$ mV	&$97.9$	&$96.2 \pm 0.1$	&$97.1 \pm 0.3$	&$0.0$	 	\\
		\cite{broome} Broome(R)\tnote{a} &$211 \pm 7$ ms	&$27.2 \pm 0.1$ mV	&$98.7$	&$96.6 \pm 0.1$	&$97.7 \pm 0.3$	&$+0.1$	  \\
	\hline
	\hline
	
\end{tabular}
\begin{tablenotes}
\item[a] Names within parenthesis are taken from initial reference to distinguish between readout performed on different quantum dots/transitions.
\end{tablenotes}
\end{threeparttable}
\end{table*}

We now apply the model to a real experiment and investigate the parameter space to maximise the readout fidelity. To demonstrate how to further improve readout fidelity we use the results from Broome(L)\cite{broome} who already achieved high fidelity measurements of electron spin states in a 2P donor dot system. In Fig.~\ref{fig:broomecompare}a we use our model, and the experimental values from Broome($L$) to plot the phase diagram of the optimal measurement fidelity by sweeping the sample rate of the charge sensor, $\Gamma_s$ and the filter cut-off frequency, $f_c$. By directly comparing the phase diagram to that in Fig.~\ref{fig:limitingcase_ele}a and examining the state distributions in Fig.~\ref{fig:broomecompare}b we can immediately determine that the slow sample rate of the charge sensor in this experiment is the main factor that limits the readout fidelity, similar to Fig.~\ref{fig:limitingcase_ele}c. The Shannon-Nyquist sampling theorem is usually assumed to set the sample rate of the charge sensor, $\Gamma_s{=}2 f_c$~\cite{shannon98} (red line in Fig.~\ref{fig:broomecompare}); however, the Shannon-Nyquist theorem only applies to signals that contain no frequency components above the filter cut-off frequency. When performing readout this is never true as the tunnel events follow an exponential distribution and there will always be some events that are above the filter frequency. Using the theory presented here, we obtain a fidelity of $97.0$~\% using $\Gamma_s{=}5$~kHz and $f_c{=}1$~kHz as in the experiment (black diamond in Fig.~\ref{fig:broomecompare}a and Fig.~\ref{fig:broomecompare}b). The readout fidelity was limited by a combination of the low filter frequency and a slightly slow sample rate. However, our analysis shows this could be improved to $F_M{=}97.9$~\% by simply using $\Gamma_s{=}5.5$~kHz and $f_c{=}2$~kHz, shown in Fig.~\ref{fig:broomecompare}c. This is a small but significant increase in fidelity that can be easily identified, demonstrating the value of the model presented here.

By using our model with parameters obtained from previous results (see Supplementary Materials III), we can compare calculated fidelities with those quoted in past work, which we show in Table~\ref{table2}. Whilst the fidelities of the model agree well with previous quoted fidelities, since the methods used to calculate the fidelities in each paper have differed, it has not been possible to make direct comparisons between them. The analysis presented here makes it possible to compare different samples for the first time and we show that by including the state-to-charge conversion, some of the previously quoted fidelities will be reduced. Also, the error in $F_M$ is lower for calculated values than for reported values (e.g. Broome(L) and Broome(R)\cite{broome}) as expected since our analytic model used for calculations removes numerical errors resultant from a Monte-Carlo simulation. Note that the parameters required to calculate the readout fidelity are often not quoted. This makes it difficult to calculate the correct readout fidelity and approximations must be used.

The limiting factor in the majority of the previous results is the electrical visibility. This is mainly due to the sensitivity index $D'$ of the charge detector used in the experiment. The state-to-charge conversion visibility is typically lower in GaAs gate defined quantum dots~\cite{Elzerman2004, Nowack} due to the comparatively lower Zeeman splitting at the same magnetic field values. Therefore, larger magnetic fields must be applied to achieve the same quality of state-to-charge conversion as compared with silicon based devices. In addition, this has the adverse effect of decreasing the electron $T_1$ relaxation times and hence, further decreases the state-to-charge conversion visibility since there is an increased chance that the spin state relaxes before being measured. Nevertheless, straight forward improvements to readout fidelities are possible for a range of previous results by optimising the values used for the readout time $t_{opt}$ and detector response threshold $x_{opt}$. We perform this optimisation for each previously reported experiment and present the results in Table~\ref{table3}. Most of the optimisations result in small, yet significant improvements to the measurement fidelity $F_M$ shown by the gain (optimised $F_M$ minus calculated $F_M$), with the most notable increasing $F_M$ by over 8\%.

\subsection{Minimisation of the readout time though optimisation of the qubit tunnel rates}

Ideally the qubit should be readout as fast as possible while still maintaining high measurement fidelity. This is particularly important when the qubits are measured sequentially and will be vital for making scalable quantum computers as fast as possible. The main limiting factor to the speed of the readout is the noise as it scales with increasing filter cut-off frequency. Therefore, we need to find the highest filter frequency where we can still perform high fidelity readout. To investigate this we need to find the dependency of $t_{\textnormal{\tiny{OUT}}}^1$ as a function of the sensitivity index  $D'$ (we assume $\sigma_0{=}\sigma_1$ for simplicity). We first calculate $f_c$ as a function of $D'$,
\begin{equation}
f_c = \frac{(\mu_1 - \mu_0)^2}{2 D'^2 A_n^2}.
\label{eq:fc_calc}
\end{equation}

The optimisation of the qubit tunnel rates that determine the $0$ and $1$ levels is difficult due to the many factors involved in the fidelity calculations. Firstly, we assume that the readout is performed at zero detuning between the spin states, such that $t_{\textnormal{\tiny{OUT}}}^1{=}t_{\textnormal{\tiny{IN}}}^0$. Secondly, $T_1$ and $t_{\textnormal{\tiny{OUT}}}^0$ are both much longer than $t_{\textnormal{\tiny{OUT}}}^1$ so that the STC visibility is not limiting the overall readout fidelity (${>}100 00t_{\textnormal{\tiny{OUT}}}^1$, corresponding to $\Delta E{\approx}18k_B T$). Finally, for simplicity we use the Shannon-Nyquist theorem to set the sample rate of the charge sensor, $\Gamma_s{=}2f_c$ despite this sample rate not being optimal.

Using the assumptions outlined in the above, we show in Fig.~\ref{fig:tuntimes}a the normalised fastest tunnel rate where $F_M{>}99$ \%, $\Gamma_{\textnormal{\tiny{OUT}}}^1{=}1/t_{\textnormal{\tiny{OUT}}}^1$ of the qubit $\ket{1}$ state for $\Delta E{\approx}18k_B T$ (red) and $\Delta E{\approx}13k_B T$ (blue) as a function of SNR ($D'$) obtained by changing the filter cut-off frequency. The latter case ($\Delta E{\approx}13k_B T$) is plotted as this is approximately the lowest qubit energy splitting where $F_M{>}99$ \% can still be achieved. We can see a clear peak in the tunnel rate where the fastest readout time can be achieved. For $\Delta E{\approx}18k_B T$ the fastest tunnel rate occurs near $D'{\approx}5.75$, while for $\Delta E{\approx}13k_B T$ the peak is slightly shifted to $D'{\approx}6$. The fact that there is an optimal $D'$ may be somewhat surprising. For low $D'$ the filter frequency is high and as a result to achieve high fidelity readout the tunnel times must be quite long compared to the filter frequency to ensure there are enough high level points in the charge sensor trace. This is to account for the lower noise, which essentially means that the $1$ state must be sampled more to obtain a high probability of a high maximum charge sensor response. At high $D'$ the filter frequency is low and as a result the tunnel time must be slow to ensure that none of the tunnel events are attenuated and occur below the charge sensor threshold. This means that although the ratio $f_c/\Gamma_{\textnormal{\tiny{OUT}}}^1$ may be the smallest for high $D'$ the tunnel time is still slower compared to the lower noise case. The optimal $D'$ occurs where these two competing effects are minimised.

\begin{figure}[t!]
\begin{center}
\includegraphics[width=1\columnwidth]{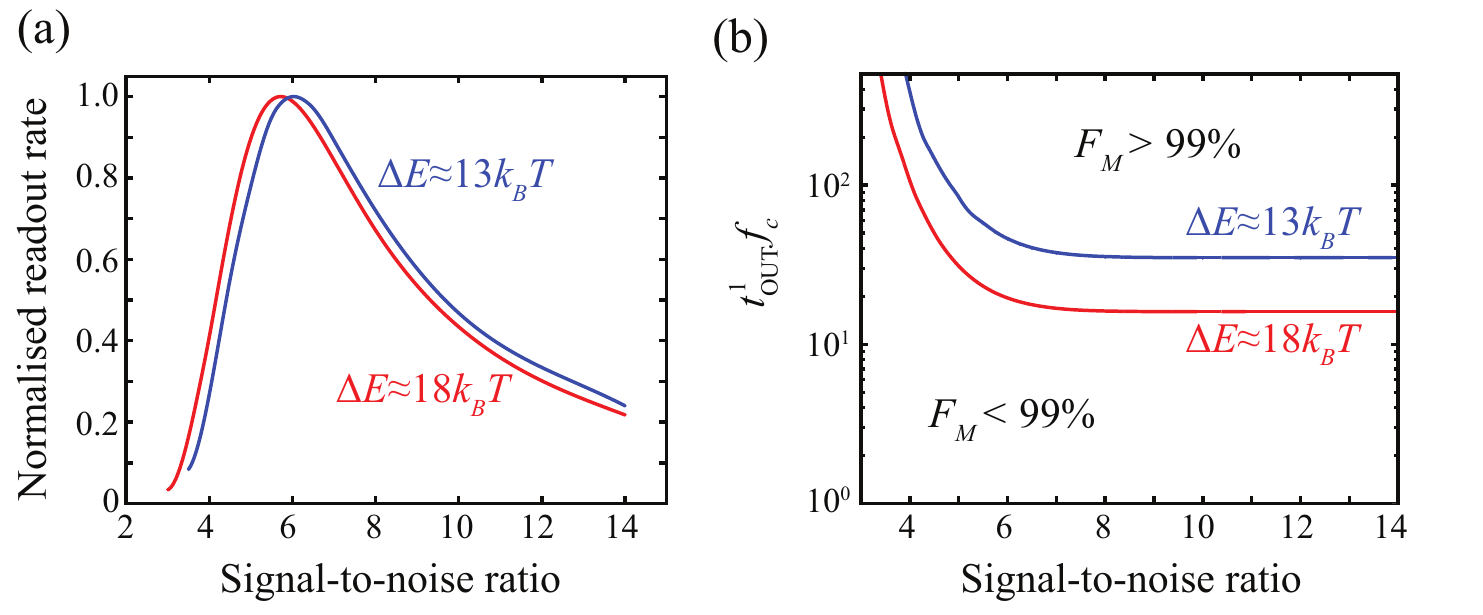}
\end{center}
\caption{{\bf Optimisation of the tunnel rate of the qubit state to the reservoir}. (a) The normalised tunnel rate (to the maximum obtainable tunnel rate) as a function of the signal-to-noise ratio for $\Delta E{\approx}18k_B T$ (red) and $\Delta E{\approx}13k_B T$ (blue) obtained by varying the filter cut-off frequency. There is a clear peak in both cases near $D'{\approx}5.75$ for $\Delta E{\approx}18k_B T$ and $D'{\approx}6$ for $\Delta E{\approx}13k_B T$. (b) The ratio of the maximum qubit tunnel rate to the filter frequency as a function of $D'$ for $\Delta E{\approx}18k_B T$ (red) and $\Delta E{\approx}13k_B T$ (blue). The line bounds the regions between $F_M{=}99$ \% and can be used as a cut-off tunnel rate between $F_M{>}99$ \% and $F_M{<}99$ \%.}
\label{fig:tuntimes}
\end{figure}

The minimum $t_{\textnormal{\tiny{OUT}}}^1$ for $F_M{=}99$ \% is $t_{\textnormal{\tiny{OUT}}}^1{=}t_{\textnormal{\tiny{IN}}}^0{\gtrsim}21/f_c$ for $\Delta E{\approx}18k_B T$ ($D'{=}5.75$) and $t_{\textnormal{\tiny{OUT}}}^1{=}t_{\textnormal{\tiny{IN}}}^0{\gtrsim}50/f_c$ for $\Delta E{\approx}13k_B T$ ($D'{=}6.00$). Note that these plots represent the minimum tunnel rate where $F_M{=}99$ \%. Therefore, as an additional investigation we also plot $t_{\textnormal{\tiny{OUT}}}^1f_c$ as a function of $D'$ to find the border between $F_M{>}99$ \% and $F_M{<}99$ \% in Fig.~\ref{fig:tuntimes}b. The region above the lines show where $F_M{>}99$ \% can be obtained for both $\Delta E{\approx}18k_B T$ (red) and $\Delta E{\approx}13k_B T$ (blue). Below the line the total measurement fidelity is always less than $99$ \%. The results here show there is a large variation of the fastest tunnel rates that can be obtained depending on the $D'$ of the charge sensor. Importantly, it appears that increasing the $D'$ above ${\sim}6$ appears to have a minimal effect on the overall measurement fidelity (demonstrated by the plateau region in Fig.~\ref{fig:tuntimes}b). Finally, we note that these plots were generated assuming white Gaussian noise and therefore may not be entirely applicable to charge sensors with, for example, $1/f$ dominated noise.

\subsection{Extension to sequential multi-qubit readout}

Single electron spin measurement has already been demonstrated over ten years ago~\cite{Elzerman2004} and the semiconductor quantum computing field is moving towards sequential multi-spin readout~\cite{Nowack,Watsone1602811}. As such, we note the only extension to the model presented here to incorporate multi-spin readout is to take into account the extra wait time while reading out the other qubit(s). Neglecting any cross-talk between the qubits during the individual qubit readout time we only need to take into account the relaxation of the $\ket{1}$ state. For the following analysis we assume that when the other qubit(s) is/are not being measured they have no probability of tunnelling out to the reservoir, that is, $t^1_{\textnormal{\tiny{OUT}}}{=}t^0_{\textnormal{\tiny{OUT}}}{\rightarrow}\infty$. Therefore, the only relevant time scale is the relaxation time of $\ket{1}$, $T_1$.

We assume that qubit $i$ is measured for a time, $t_{m,i}$ which can be optimised independently using Eq.~\ref{eqn:topt} and that the total readout time for all qubits is $T_m{=}\sum_i t_{m,i}$. In addition, we define the total time \emph{before} qubit $i$ is measured as the wait time for qubit $i$,
\begin{equation}
t_{w,i} = \sum_{j=1}^{i-1} t_{m,j}, \forall i > 1, \hspace{0.5cm} \textnormal{and} \hspace{0.5cm} t_{w,1} = 0.
\end{equation}

The only modification to account for the extended wait time is a multiplicative factor in Eq.~\ref{eqn:fstc1} that accounts for the probability that $\ket{1}$ relaxes during the measurement of the other qubits,
\begin{equation}
F_{STC,i}^1 = \Lambda_i F_{STC}^1(t^0_{\textnormal{\tiny{OUT}},i},t^1_{\textnormal{\tiny{OUT}},i},T_{1,i})
\label{eqn:fstctw}
\end{equation}
where $\Lambda_i{=}\textnormal{exp}(-t_{w,i}/T_{1,i})$, $T_{1,i}$ is the relaxation time of qubit $i$ and $F_{STC}^1$ is defined in Eq.~\ref{eqn:fstc1}. Equation~\ref{eqn:fstctw} can then be used instead of Eq.~\ref{eqn:fstc1} for calculation of $F_M$. Note that $t_{opt}$ obtained using Eq.~\ref{eqn:topt} will still give the optimal values for Eq.~\ref{eqn:fstctw}. We now want to find the optimal ordering of the qubits to achieve the highest fidelity readout across all those being measured. We will demonstrate this with an example. We want to measure three different qubits, $\{Q_1, Q_2, Q_3\}$ sequentially with the following values for $t_m$ and $T_1$:

\begin{table}[h]
\centering
\begin{tabular}{ c | c  c  c }
& $Q_1$ & $Q_2$ & $Q_3$\\
\hline
$t_m$ & 3 & 1 & 2\\
$T_1$ & 5 & 2 & 10
\end{tabular}
\end{table}

Since we want to maximise the fidelity across all qubits, we are only interested in the multiplicative factor in Eq.~\ref{eqn:fstctw}. Therefore, we need to calculate $\Lambda_i$ for every order of measurement, $M{=}\{M_1, M_2, M_3\}$. This means there are $N!$ measurement combinations we must consider where $N$ is the total number of qubits to be read out. Whichever qubit is measured first, by definition, has $\Lambda_1{=}1$ since $t_{w,1}{=}0$. To determine the best measurement order we calculate $\sum \Lambda_i/N$ which corresponds to the average reduction in $F_{STC}^1$ across all the qubits. As $\sum \Lambda_i/N{\rightarrow}1$ the higher the overall fidelity will be obtained using the given measurement order. From the calculations in the table below we can see that the measurement $\{M_1, M_2, M_3\}{=}\{Q_2, Q_1, Q_3\}$ will be optimal for mitigating the effect of sequential readout on the individual qubits since this combination has the largest $\sum \Lambda_i/N$.

\begin{table}[h]
\centering
\begin{tabular}{ c  c  c | c  c  c | c}
$M_1$ & $M_2$ & $M_3$ & $\Lambda_1$ & $\Lambda_2$ & $\Lambda_3$ & $\sum \Lambda_i/N$\\
\hline
$Q_1$ & $Q_2$ & $Q_3$ & 1.0000 & 0.2231 & 0.6703 & 0.6311\\
$Q_1$ & $Q_3$ & $Q_2$ & 1.0000 & 0.7408 & 0.0821 & 0.6076\\
$Q_2$ & $Q_1$ & $Q_3$ & 1.0000 & 0.8187 & 0.6703 & 0.8297\\
$Q_2$ & $Q_3$ & $Q_1$ & 1.0000 & 0.9048 & 0.5488 & 0.8179\\
$Q_3$ & $Q_1$ & $Q_2$ & 1.0000 & 0.6703 & 0.0821 & 0.5841\\
$Q_3$ & $Q_2$ & $Q_1$ & 1.0000 & 0.3679 & 0.5488 & 0.6389\\
\end{tabular}
\end{table}

We can immediately see here that the average $F_{STC}^1$ across all qubits is reduced by a factor of $\sum \Lambda_i/N{=}0.8297$ compared to measuring each qubit simultaneously, that is, without waiting between measurements. The method outlined above can also be used to optimise the measurement order for the readout of specific qubits.

\section{Summary}

Single-shot electron spin readout fidelity calculations will become increasingly important as experiments push towards the fault-tolerant threshold for 2-dimensional surface codes~\cite{PhysRevLett.107.146801,shulman2012,Veldhorst2015}. The improvement of gate fidelities will place increasingly more emphasis on state preparation and measurement errors as these become the limiting source of infidelity. The current state of analysis for measurement fidelity varies considerably~\cite{Elzerman2004,Morello2010,Buch2013} and in this current paper we propose a standard approach, which we have used to make a comparison between previous experimental results.

We have presented a method to calculate the single-shot readout fidelity of a detector based on an a comprehensive statistical analysis of the system. We first provided a simple formula to calculate the sample rate and readout time required to achieve high fidelity readout, where we emphasise the importance of choosing a sufficiently fast sample rate. Using our model, we describe different fidelity limiting factors, how to identify them from the model and strategies to increase the fidelity once they have been identified. To illustrate this we use the results from a 2P donor dot system that had already achieved high fidelity electron spin readout (Broome($L$))\cite{broome} to show that the fidelity can be further increased by 0.9~\% since the previous measurements had been limited by the sample rate and filter frequency. Assuming white Gaussian noise and provided the key experimental parameters are met: charge sensor sample rate $\Gamma_s{\gtrsim}12{/}t^0_{\textnormal{\tiny{IN}}}$, $D'{\gtrsim}3$, qubit energy splitting $\Delta E{\gtrsim}13k_B T$ and a long relaxation time, $T_1{\gtrsim}100t^1_{\textnormal{\tiny{OUT}}}$ then fidelities greater than 99~\% can be achieved.

\begin{acknowledgements}
The research was conducted by the Australian Research Council Centre of Excellence for Quantum Computation and Communication Technology (project number CE170100012). MYS acknowledges an Australian Research Council Laureate Fellowship.
\end{acknowledgements}


\setcounter{figure}{0}
\setcounter{section}{0}
\setcounter{table}{0}
\setcounter{equation}{0}
\makeatletter 
\renewcommand{\thefigure}{S\arabic{figure}}
\renewcommand{\theequation}{S\arabic{equation}}
\renewcommand{\thetable}{S\arabic{table}}
\makeatother

\begin{widetext}
\vspace{10cm}

\newpage

\section*{\Large{SUPPLEMENTARY MATERIALS}}

\section{Comparison of the numerical approach and the model}
Numerical methods are inherently less accurate than analytical solutions. There is an associated error in the Monte-Carlo methods previously used that is often not quoted in the fidelity calculations. For example, in Fig. ~\ref{fig:dists}a we plot the calculated fidelity using the numerical approach in~\cite{Watson2015} as a function of the number of simulation runs used to bin the fidelity histograms. We repeat this 100 times for each number of simulations while keeping the bin size of the histogram the same. We can see that even with 500 000 simulation runs and 1000 bins there is still an error of 0.2~\%.

We now want to compare the analytical calculation of the electrical visibility with the Monte-Carlo method. For this, we use the Watson $D^{-}$\cite{Watson2015} data. In figure~\ref{fig:dists}b we plot the numerically evaluated histogram and the one obtained using our model. We emphasise that analytical model is \textit{not} a fit but is calculated using the same set of parameters in Tab.~\ref{table1}.

\begin{figure}
\begin{center}
\includegraphics[width=1\columnwidth]{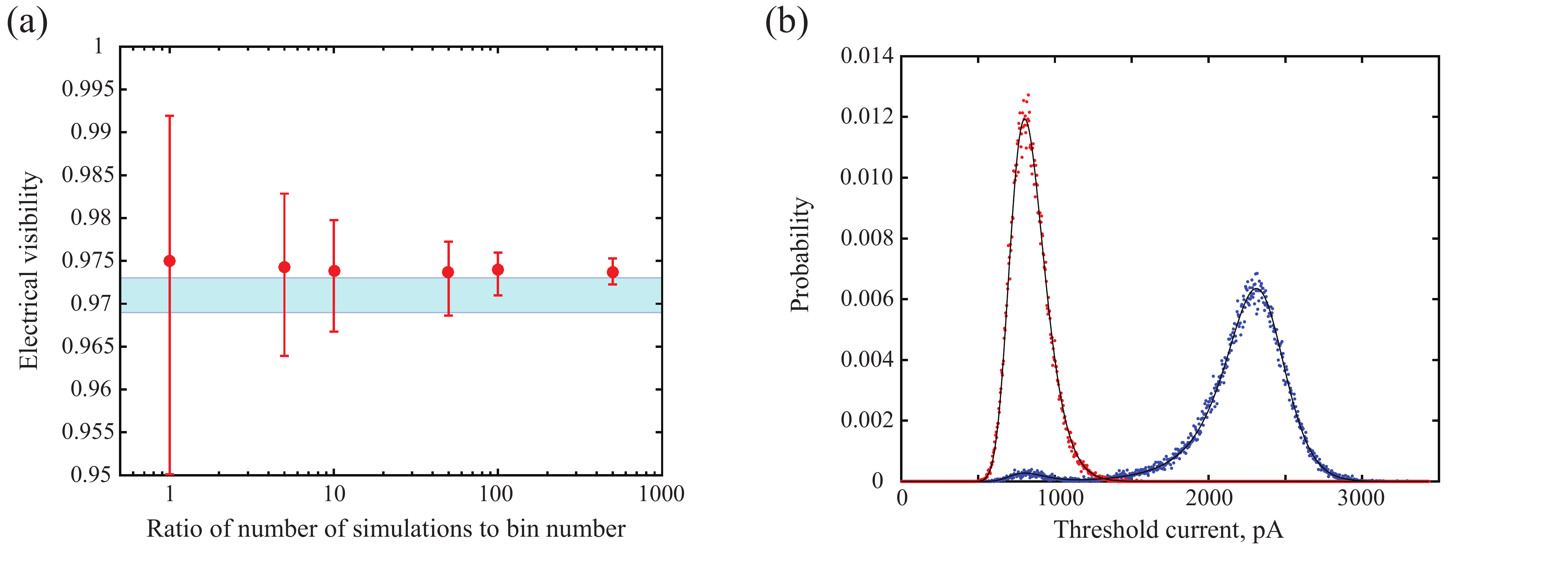}
\end{center}
\caption{{\bf Comparison of the Monte-Carlo method and the model presented here}. a) The calculated electrical visibility with associated error as a function of the number of simulations used to generate the readout distributions using the Monte-Carlo method. The shaded blue area represents the value calculated with the model with errors obtained from the experimental parameters. b) The readout distributions, $P_E^0$ ($P_E^1$) calculated using the Monte-Carlo method shown by the red (blue) circles and the model presented in this paper (black lines). The numerical histograms are obtained using 1000 bins and 100 000 simulation traces. Both sets of data use the same input parameters and no fitting is performed.} 
\label{fig:dists}
\end{figure}

\section{Parameters for a 8th order low pass Bessel filter}

Here we give the relevant parameters for the Bessel filter. A filter can be characterised by its transfer function, $\mathcal{T}(f)$. For a Bessel filter, $\mathcal{T}(f)$ is made of reverse Bessel polynomials \cite{Grosswald1978} of the same order as the filter. That is, for a 8th order filter,
\begin{equation}
\mathcal{T}(f) = \frac{\theta_8(0)}{\theta_8(f)}
\end{equation}
where $\theta_8(f)$ is the $8$th order reverse Bessel polynomial and the gain function can be obtained,
\begin{equation}
\mathcal{G}(f,f_c) = |\mathcal{T}\Big(i \frac{f}{f_c}\Big)|
\end{equation}
Therefore, if $\mathcal{T}(f)$ is known for a particular filter, then $\mathcal{G}(f,f_c)$ can be easily calculated. The pulse response for a filter can be obtained from the inverse Laplace transform of the transfer function,
\begin{equation}
h_p(t,f_c) = \mathcal{L}^{-1}\{\mathcal{T}\Big(\frac{f}{f_c}\Big) \Big(\frac{1 - e^{- t f}}{f}\Big)\}
\end{equation}
where the pulse length in time is $t$. For low order filters, this can be found analytically. However, for the higher order Bessel filters it is more convenient to use the gain function with a small correction due to overshoot,
\begin{equation}
\max{[h_p(n,m_c)]} \approx O(q) \mathcal{G}(f,f_c),
\end{equation}
where $q$ is the order of the filter, $O(q){=}A_{max}{+}1$ is the overshoot correction and $A_{max}$ is the percentage overshoot of the filter. To find the approximate overshoot factor of the Bessel filter we numerically simulated a single `blip' of height 1 in the readout trace (without noise) which we then filter and calculated the maximum of the trace. We varied the frequency of the blip and arrived at $A_{max}{\approx}0.00344$, hence we find $O(8){\approx}1.00344$.

Finally, the introduction of a filter also introduces frequency dependent noise. As a result, the noise on the readout becomes slightly correlated. The amount of correlation is dependent on the sample rate of the readout trace, $\Gamma_s$ compared to the filter cut-off frequency, $f_c$. To account for the correlated points, we introduce a normalised frequency, $f_s{=}2 f_c {/} \Gamma_s$. The correlation between the readout points means they are not independent and we cannot take the $n_r^{\text{th}}$ power in Eq.~51, 54, and 55. Instead, $n_r$ is given by,
\begin{equation}
	n_r \rightarrow
	\begin{cases}
		\frac{2 f_s}{f_s + 1} n_r & f_s < 1, \\
		n_r & \text{otherwise},
	\end{cases}
\end{equation}
which accounts for the correlation introduced from the filter. Finally, the effective blip length in $P_{miss}$ also needs to be extended due to the correlation from the filter. This is taken into account by a similar transformation,
\begin{equation}
	R^i_s \rightarrow
	\begin{cases}
		\frac{2 f_s}{f_s + 1} R^i_s & f_s < 1, \\
		R^i_s & \text{otherwise}.
	\end{cases}
\end{equation}

\newpage

\section{Parameters used in the calculation of previous experimental fidelities}

Table~\ref{table1} contains the quoted, assumed or calculated parameters required to fully characterise the single-shot readout fidelity.

\begin{table*}[h!]
\begin{threeparttable}
\caption{Parameters used in the determination of the single-shot readout fidelities and optimal readout time and threshold value. $\mu_i$ is the detector level of the charge sensor, $A_n$ is the noise power spectral density of the charge sensor, $f_c$ is the filter cut-off frequency, $\Gamma_s$ is the sample rate of the charge sensor, $t^1_{\textnormal{OUT}}$ is the tunnel out time of $\ket{1}$, $t^0_{\textnormal{OUT}}$ is the tunnel out time of $\ket{0}$, $t^0_{\textnormal{IN}}$ is the tunnel in time of $\ket{0}$. The $B$ values are the magnetic fields for which the given $T_1$ times were measured.}
\label{table1}
\centering
\begin{tabular}{ l | C{1.4cm}  C{2.3cm}  C{1.3cm}  C{1.5cm}  c  C{1.5cm}  C{1.5cm}  C{1.3cm}  C{0.8cm}  c }
	\hline
	\hline
	Reference 		& $\mu_1 - \mu_0$ 	& $A_n$ 								& $f_c$[kHz] 		& $\Gamma_s$ 	& $t^1_{\textnormal{OUT}}$ 		& $t^0_{\textnormal{OUT}}$ 		& $t^0_{\textnormal{IN}}$ 			& $T_1$					& $B$[T] & $T$[mK] \\
	\hline
		\cite{Elzerman2004} Elzerman  			& $0.37$ nA 		& $0.31$ pA/$\sqrt{\textnormal{Hz}}$\tnote{a} 	& $40$  	& $2f_c$\tnote{a}  	& $0.11$ ms 			& $13.6$ ms\tnote{b} 			& $0.11$ ms\tnote{b} 			& $0.55$ ms		& $10$ & $300$ \tnote{f} \\
		\cite{Morello2010} Morello  			& $1.9$ nA 		& $0.80$ pA/$\sqrt{\textnormal{Hz}}$\tnote{a}		& $120$ 	& $2f_c$\tnote{a} 		& $10$ $\mu$s 		& $392.7$ s\tnote{b}			& $40$ $\mu$s 		& $\sim 0.03$ s			& $5$ & $200$\\
		\cite{PhysRevLett.106.156804} Simmons  	& $15.5$ pA\tnote{a} 		& $28$ fA/$\sqrt{\textnormal{Hz}}$\tnote{a} 	& $20$\tnote{a}  	& $40$ kHz  	& $21.6$ ms 	& $13.8$ s\tnote{b} 		& $9.4$ ms 	& $2.72$ s	& $1.85$ & $143$ \\
		\cite{Nowack} Nowack(R)\tnote{i}  		& $300$ pA\tnote{c} & $0.17$ pA/$\sqrt{\textnormal{Hz}}$\tnote{a} 	& $40$ \tnote{f} 	& $2f_c$\tnote{a}  	& $0.5$ ms\tnote{c} 			& $16.6$ ms\tnote{b} 			& $0.3$ ms\tnote{b} 			& $3.8$ ms			& $6.5$ & $250$\\
		\cite{Pla2012} Pla 				& $1.1$ nA\tnote{c} & $0.16$ pA/$\sqrt{\textnormal{Hz}}$\tnote{a} 	& $120$ \tnote{a} 	& $2f_c$\tnote{a} 	& $295$ $\mu$s 		& $1.2$ ms\tnote{b} 			& $33$ $\mu$s 		& $6$ s\tnote{g}					& $1.07$ & $300$ \\
		\cite{Buch2013} Buch  			& $32$ pA 		& $71$ fA/$\sqrt{\textnormal{Hz}}$ 	& $0.3$	& $1.2$ kHz\tnote{a}	 	& $3.9$ ms 			& $1070$ ms 			& $6.9$ ms 			& $1.85$ s				& $1.2$ & $200$ \\
		\cite{VeldhorstM.2014} Veldhorst  		& $400$ pA\tnote{h}  & $0.35$ pA/$\sqrt{\textnormal{Hz}}$\tnote{h} 	& $50$ \tnote{a} 	& $2f_c$\tnote{a}	& $22$ $\mu$s 		& $22.8$ ms\tnote{b} 			& $127$ $\mu$s 		& $1$ s\tnote{d}					& $1.4$  & $150$ \tnote{e}\\
		\cite{Watson2015} Watson($D^0$)\tnote{i}  	& $450$ pA		& $0.24$ pA/$\sqrt{\textnormal{Hz}}$ 	& $10$  	& $2f_c$  	& $6.5$ ms 	& $24$s 		& $5.1$ ms 	& $4$ s					& $1.6$ & $160$\\
		\cite{Watson2015} Watson($D^-$)\tnote{i}  	& $1.72$ nA 		& $0.69$ pA/$\sqrt{\textnormal{Hz}}$ 	& $100$  	& $2f_c$  	& $0.14$ ms 	& $145$ ms 		& $0.13$ ms 	& $4$ s					& $1.6$ & $160$ \\
		\cite{Watsone1602811} Watson($D1$)\tnote{i} 	& $153$ pA		& $0.2$ pA/$\sqrt{\textnormal{Hz}}$ 	& $10$  	& $2f_c$  	& $5.7$ ms 	& $162$ s 		& $11.6$ ms 	& $30$ s					& $1.5$ & $100$\\
		\cite{Watsone1602811} Watson($D2$)\tnote{i}  	& $153$ pA 		& $0.2$ pA/$\sqrt{\textnormal{Hz}}$ 	& $10$  	& $2f_c$  	& $5.9$ ms 	& $99$ s 		& $8.2$ ms 	& $15$ s					& $1.5$ & $100$ \\
		\cite{broome} Broome(L)\tnote{i} 		& $50$ mV 		& $133$ $\mu$V/$\sqrt{\textnormal{Hz}}$ 	& $1$  	& $5f_c$  	& $1.83$ ms 	& $0.61$ s 	& $6.62$ ms 	& $2.9$ s			& $2.5$ & $200$ \\
		\cite{broome} Broome(R)\tnote{i}		& $49$ mV 		& $133$ $\mu$V/$\sqrt{\textnormal{Hz}}$ 	& $1$  	& $5f_c$  	& $31.6$ ms 	& $25$ s 		& $7.65$ ms 	& $9.3$ s			& $2.5$ & $200$ \\
	\hline
	\hline
	
\end{tabular}
\begin{tablenotes}
\item[a] Value was estimated by matching the current histogram, shown in the paper, with our analytical model.
\item[b] Value approximated using Eq. 8 and 9.
\item[c] Due to incomplete data available, value was estimated from the peak current. distribution shown in a given reference
\item[d] Estimation based on similar work \cite{yang2013}
\item[e] Estimation based on similar work \cite{Lai}
\item[f] Estimation based on similar work \cite{vandersypen2004}
\item[g] Estimation based on similar work \cite{Morello2010}
\item[h] Inconsistent data. Estimation based on main text.
\item[i] Names within parenthesis are taken from initial reference to distinguish between readout performed on different quantum dots/transitions.
\end{tablenotes}
\end{threeparttable}
\end{table*}

\end{widetext}


\begin{thebibliography}{1}

\bibitem{DiVincenzo2000}
D DiVincenzo
\newblock { The Physical Implementation of Quantum Computation}.
\newblock {\em Fortschritte der Physik}, 48:771--783, 2000.

\bibitem{PhysRevA.83.020302}
Wang, David S. and Fowler, Austin G. and Hollenberg, Lloyd C. L.
\newblock { Surface code quantum computing with error rates over 1\%}.
\newblock {\em Phys. Rev. A}, 83:020302, 2011.

\bibitem{Fowler2012}
Fowler, Austin G. and Mariantoni, Matteo and Martinis, John M. and Cleland, Andrew N.
\newblock { Surface codes: Towards practical large-scale quantum computation}.
\newblock {\em Phys. Rev. A}, 86:032324, 2012.

\bibitem{Hille1500707}
Hill, Charles D. and Peretz, Eldad and Hile, Samuel J. and House, Matthew G. and Fuechsle, Martin and Rogge, Sven and Simmons, Michelle Y. and Hollenberg, Lloyd C. L.
\newblock { A surface code quantum computer in silicon}.
\newblock {\em Science Advances}, 1:e1500707, 2015.

\bibitem{ogorman2016}
J. O'Gorman and N. H. Nickerson and P. Ross and J. J. L. Morton and S. C. Benjamin
\newblock { A silicon-based surface code quantum computer}.
\newblock {\em npj Quant. Infor.}, 2:15019, 2016.

\bibitem{PhysRevA.77.012307}
Knill, E. and Leibfried, D. and Reichle, R. and Britton, J. and Blakestad, R. B. and Jost, J. D. and Langer, C. and Ozeri, R. and Seidelin, S. and Wineland, D. J.
\newblock { Randomized benchmarking of quantum gates}.
\newblock {\em Phys. Rev. A}, 77:012307, 2008.

\bibitem{PhysRevLett.106.180504}
Magesan, Easwar and Gambetta, J. M. and Emerson, Joseph
\newblock { Scalable and Robust Randomized Benchmarking of Quantum Processes}.
\newblock {\em Phys. Rev. Lett.}, 106:180504, 2011.

\bibitem{Barends2014}
Barends, R. and Kelly, J. and Megrant, A. and Veitia, A. and Sank, D. and Jeffrey, E. and White, T. C. and Mutus, J. and Fowler, A. G. and Campbell, B. and Chen, Y. and Chen, Z. and Chiaro, B. and Dunsworth, A. and Neill, C. and O'Malley, P. and Roushan, P. and Vainsencher, A. and Wenner, J. and Korotkov, A. N. and Cleland, A. N. and Martinis, John M.\newblock { Superconducting quantum circuits at the surface code threshold for fault tolerance}.
\newblock {\em Nature}, 508:500--503, 2014.

\bibitem{1367-2630-16-10-103032}
Joel J Wallman and Steven T Flammia
\newblock { Randomized benchmarking with confidence}.
\newblock {\em New Journal of Physics}, 16:103032, 2014.

\bibitem{Cross2016}
Cross, Andrew W. and Magesan, Easwar and Bishop, Lev S. and Smolin, John A. and Gambetta, Jay M.
\newblock { Scalable randomised benchmarking of non-Clifford gates}.
\newblock {\em npj Quant. Infor.}, 2:16012, 2016.

\bibitem{PhysRevLett.113.220501}
Harty, T. P. and Allcock, D. T. C. and Ballance, C. J. and Guidoni, L. and Janacek, H. A. and Linke, N. M. and Stacey, D. N. and Lucas, D. M. 
\newblock { High-Fidelity Preparation, Gates, Memory, and Readout of a Trapped-Ion Quantum Bit}.
\newblock {\em Phys. Rev. Lett.}, 113:220501, 2014.

\bibitem{loss1998}
D. Loss and D. P. DiVincenzo
\newblock { Quantum computation with quantum dots}.
\newblock {\em Phys. Rev. A}, 57:120--126, 1998.

\bibitem{kane1998}
B. E. Kane
\newblock { A silicon-based nuclear spin quantum computer}.
\newblock {\em Nature}, 393:133--137, 1998.

\bibitem{Tosi2017}
Tosi, Guilherme and Mohiyaddin, Fahd A. and Schmitt, Vivien and Tenberg, Stefanie and Rahman, Rajib and Klimeck, Gerhard and Morello, Andrea
\newblock { Silicon quantum processor with robust long-distance qubit couplings}.
\newblock {\em Nature Communications}, 8:450, 2017.

\bibitem{Veldhorst2017}
Veldhorst, M. and Eenink, H. G. J. and Yang, C. H. and Dzurak, A. S.
\newblock { Silicon CMOS architecture for a spin-based quantum computer}.
\newblock {\em Nature Communications}, 8:1766, 2017.

\bibitem{Elzerman2004}
Elzerman, J. M. and Hanson, R. and Willems van Beveren, L. H. and Witkamp, B. and Vandersypen, L. M. K. and Kouwenhoven, L. P.
\newblock { Single-shot read-out of an individual electron spin in a quantum dot}.
\newblock {\em Nature}, 431--435, 2004.

\bibitem{Morello2010}
Morello, Andrea and Pla, Jarryd J. and Zwanenburg, Floris A. and Chan, Kok W. and Tan, Kuan Y. and Huebl, Hans and Mottonen, Mikko and Nugroho, Christopher D. and Yang, Changyi and {van Donkelaar}, Jessica A. and Alves, Andrew D. C. and Jamieson, David N. and Escott, Christopher C. and Hollenberg, Lloyd C. L. and Clark, Robert G. and Dzurak, Andrew S.
\newblock { Single-shot readout of an electron spin in silicon}.
\newblock {\em Nature}, 467:687--691, 2010.

\bibitem{Buch2013}
B{\"u}ch,H. and Mahapatra,S. and Rahman,R. and Morello,A. and Simmons,M. Y.
\newblock { Spin readout and addressability of phosphorus-donor clusters in silicon}.
\newblock {\em Nature Communications}, 4:2017, 2013.

\bibitem{PhysRevLett.106.156804}
Simmons, C. B. and Prance, J. R. and Van Bael, B. J. and Koh, Teck Seng and Shi, Zhan and Savage, D. E. and Lagally, M. G. and Joynt, R. and Friesen, Mark and Coppersmith, S. N. and Eriksson, M. A.
\newblock { Tunable Spin Loading and ${T}_{1}$ of a Silicon Spin Qubit Measured by Single-Shot Readout}.
\newblock {\em Phys. Rev. Lett.}, 106:156804, 2011.

\bibitem{VeldhorstM.2014}
Veldhorst, M. and Hwang, J. C., C. and Yang, C., H. and Leenstra, A. W. and de Ronde, B. and Dehollain, J. P. and Muhonen, J. T. and Hudson, F. E. and Itoh, K. M. and Morello, A. and Dzurak, A. S.
\newblock { An addressable quantum dot qubit with fault-tolerant control-fidelity}.
\newblock {\em Nature Nanotechnology}, 9:981--985, 2014.

\bibitem{Robledo2011}
Robledo, L. and Childress, L. and Bernien, H. and Hensen, B. and Alkemade, P. F. A. and Hanson, R.
\newblock { High-fidelity projective read-out of a solid-state spin quantum register}.
\newblock {\em Nature}, 477:574--578, 2011.

\bibitem{Watson2015}
Watson, T. F. and Weber, B. and House, M. G. and B\"uch, H. and Simmons, M. Y.
\newblock { High-Fidelity Rapid Initialization and Read-Out of an Electron Spin via the Single Donor ${D}^{-}$ Charge State}.
\newblock {\em Phys. Rev. Lett.}, 115:166806, 2015.

\bibitem{PhysRevA.89.012313}
D'Anjou, B. and Coish, W. A.
\newblock { Optimal post-processing for a generic single-shot qubit readout}.
\newblock {\em Phys. Rev. A}, 89:012313, 2014.

\bibitem{Pla2012}
Pla, J. J. and Tan, K. Y. and Dehollain, J. P. and Lim, W. H. and Jamieson, D. N. and Dzurak, A. S. and Morello, A
\newblock { A single-atom electron spin qubit in silicon}.
\newblock {\em Nature}, 489:541--545, 2012.

\bibitem{Watsone1602811}
Watson, Thomas F. and Weber, Bent and Hsueh, Yu-Ling and Hollenberg, Lloyd C. L. and Rahman, Rajib and Simmons, Michelle Y.
\newblock { Atomically engineered electron spin lifetimes of 30 s in silicon}.
\newblock {\em Science Advances}, 3:3, 2017.

\bibitem{PhysRevApplied.8.034019}
Gorman, S. K. and He, Y. and House, M. G. and Keizer, J. G. and Keith, D. and Fricke, L. and Hile, S. J. and Broome, M. A. and Simmons, M. Y.
\newblock { Tunneling Statistics for Analysis of Spin-Readout Fidelity}.
\newblock {\em Phys. Rev. Applied}, 8:034019, 2017.

\bibitem{0957-4484-26-21-215201}
J R Prance and B J Van Bael and C B Simmons and D E Savage and M G Lagally and Mark Friesen and S N Coppersmith and M A Eriksson
\newblock { Identifying single electron charge sensor events using wavelet edge detection}.
\newblock {\em Nanotechnology}, 26:215201, 2015.

\bibitem{PhysRevLett.111.126803}
House, M. G. and Xiao, Ming and Guo, GuoPing and Li, HaiOu and Cao, Gang and Rosenthal, M. M. and Jiang, HongWen
\newblock { Detection and Measurement of Spin-Dependent Dynamics in Random Telegraph Signals}.
\newblock {\em Phys. Rev. Lett.}, 111:126803, 2013.

\bibitem{PhysRevA.76.012325}
Gambetta, Jay and Braff, W. A. and Wallraff, A. and Girvin, S. M. and Schoelkopf, R. J.
\newblock { Protocols for optimal readout of qubits using a continuous quantum nondemolition measurement}.
\newblock {\em Phys. Rev. A}, 76:012325, 2007.

\bibitem{HoEomDayLeDucEtAl2012}
Ho Eom, Byeong and Day, Peter K. and LeDuc, Henry G. and Zmuidzinas, Jonas
\newblock { A wideband, low-noise superconducting amplifier with high dynamic range}.
\newblock {\em Nat. Phys.}, 8:623--627, 2012.

\bibitem{doi:10.1063/1.4941421}
L. A. Tracy and D. R. Luhman and S. M. Carr and N. C. Bishop and G. A. Ten Eyck and T. Pluym and J. R. Wendt and M. P. Lilly and M. S. Carroll
\newblock { Single shot spin readout using a cryogenic high-electron-mobility transistor amplifier at sub-Kelvin temperatures}.
\newblock {\em Applied Physics Letters}, 108:063101, 2016.

\bibitem{broome}
Broome, M. A. and Gorman, S. K. and House, M. G. and Hile, S. J. and Keizer, J. G. and Keith, D. and Hill, C. D. and Watson, T. F. and Baker, W. J. and Hollenberg, L. C. L. and Simmons, M. Y.
\newblock { Two-electron spin correlations in precision placed donors in silicon}.
\newblock {\em Nature Communications}, 9:980, 2018.

\bibitem{shannon98}
Shannon, C. E.
\newblock { Communication in the presence of noise}.
\newblock {\em Proc. IEEE}, 86:447--457, 1998.

\bibitem{Nowack}
Nowack, K. C. and Shafiei, M. and Laforest, M. and Prawiroatmodjo, G. E. D. K. and Schreiber, L. R. and Reichl, C. and Wegscheider, W. and Vandersypen, L. M. K.
\newblock { Single-Shot Correlations and Two-Qubit Gate of Solid-State Spins}.
\newblock {\em Science}, 333:1269--1272, 2011.

\bibitem{PhysRevLett.107.146801}
Brunner, R. and Shin, Y.-S. and Obata, T. and Pioro-Ladri\`ere, M. and Kubo, T. and Yoshida, K. and Taniyama, T. and Tokura, Y. and Tarucha, S.
\newblock { Two-Qubit Gate of Combined Single-Spin Rotation and Interdot Spin Exchange in a Double Quantum Dot}.
\newblock {\em Phys. Rev. Lett.}, 107:146801, 2011.

\bibitem{shulman2012}
M. D. Shulman and O. E. Dial and S. P. Harvey and H. Bluhm and V. Umansky and A. Yacoby
\newblock { Demonstration of entanglement of electrostatically coupled singlet-triplet qubits}.
\newblock {\em Science}, 336:202--205, 2012.

\bibitem{Veldhorst2015}
Veldhorst, M. and Yang, C. H. and Hwang, J. C. C. and Huang, W. and Dehollain, J. P. and Muhonen, J. T. and Simmons, S. and Laucht, A. and Hudson, F. E. and Itoh, K. M. and Morello, A. and Dzurak, A. S.
\newblock { A two-qubit logic gate in silicon}.
\newblock {\em Nature}, 526:410--414, 2015.

\bibitem{Grosswald1978}
E. Grosswald
\newblock{ Bessel Polynomials}.
\newblock{ \em Lecture Notes in Mathematics}, 698, 1978

\bibitem{yang2013}
Yang, CH and Rossi, A and Ruskov, R and Lai, NS and Mohiyaddin, FA and Lee, S and Tahan, C and Klimeck, Gerhard and Morello, A and Dzurak, AS
\newblock { Spin-valley lifetimes in a silicon quantum dot with tunable valley splitting}.
\newblock {\em Nature communications}, 4, 2013.

\bibitem{Lai}
Lai, NS and Lim, WH and Yang, CH and Zwanenburg, FA and Coish, WA and Qassemi, F and Morello, A and Dzurak, AS
\newblock { Pauli spin blockade in a highly tunable silicon double quantum dot}.
\newblock {\em Scientific reports}, 1, 2011.

\bibitem{vandersypen2004}
Vandersypen, LMK and Elzerman, JM and Schouten, RN and Willems van Beveren, LH and Hanson, R and Kouwenhoven, LP
\newblock { Real-time detection of single-electron tunneling using a quantum point contact}.
\newblock {\em Applied Physics Letters}, 85:4394--4396, 2004.

\end{thebibliography}
\end{document}